\pdfoutput=1
\documentclass[aps,prb,reprint,showpacs,superscriptaddress]{revtex4-1}

\usepackage{amsfonts}
\usepackage{enumerate}
\usepackage{longtable}
\usepackage{dcolumn}
\usepackage{graphicx}
\usepackage{amsmath}
\usepackage{amssymb}
\usepackage{dcolumn}
\usepackage{dsfont}
\usepackage{latexsym}
\usepackage{rotating}
\usepackage{color}
\usepackage{latexsym}
\usepackage{bbm}
\usepackage{subfigure}
\usepackage{float}
\usepackage{epsfig}
\usepackage{psfrag}
\usepackage{natbib}
\usepackage{bm}
\usepackage{amsthm}
\usepackage{eucal}
\usepackage{mathrsfs}
\usepackage{url}

\usepackage{color} 

\usepackage{hyperref}
\hypersetup{
colorlinks=true,final=true,
        linkcolor=red,
        citecolor=blue,
        filecolor=blue,
        urlcolor=blue,
}

\begin{document}

\newcommand{\be}{\begin{equation}}
\newcommand{\ee}{\end{equation}}
\newcommand{\bn}{\begin{eqnarray}}
\newcommand{\en}{\end{eqnarray}}

\title{Charge-Density-Wave Order in 2H-NbSe$_{2}$}
\author{S. Koley}
\email{skoley@phy.iitkgp.ernet.in}
\affiliation{Department of Physics, Indian Institute of Technology Kharagpur, W.B. 721302, India}
\author{N. Mohanta}
\affiliation{Department of Physics, Indian Institute of Technology Kharagpur, W.B. 721302, India}
\author{A. Taraphder}
\email{arghya@phy.iitkgp.ernet.in}
\affiliation{Department of Physics, Indian Institute of Technology Kharagpur, W.B. 721302, India}
\affiliation{Centre for Theoretical Studies, Indian Institute of Technology Kharagpur, W.B. 721302, India}

\begin{abstract}
Competition  between collective  states like  charge density  wave and
superconductivity  is  played out  in  some  of  the transition  metal
dichalcogenides unencumbered by the  spin degrees of freedom. Although
2H-NbSe$_2$ has  received much less  attention than some of  the other
members  of the family  (like 1T-TiSe$_2$  and 2H-TaSe$_2$),  it shows
superconductivity at  7.2 K and  incommensurate charge ordering  at 33
K.   Recent   experiments,   notably  Angle   Resolved   Photoemission
spectroscopy,  have cast  serious  doubts on  the  mechanism of  Fermi
surface nesting via electron-phonon interaction.  The normal state has
been  found  to  be  a  poor, incoherent  metal  and  remarkably,  the
coherence  increases in the  broken symmetry  state. From  a preformed
excitonic  liquid  scenario,  we  show  that there  exists  a  natural
understanding  of  the  experimental  data  on  2H-NbSe$_2$  based  on
electron-electron interaction.  The collective instabilities,  in this
scenario,  are viewed  as a  condensation of  an  incoherent excitonic
liquid already present at high temperature.
\end{abstract}
\maketitle

\section{Introduction}
Strongly  correlated electronic systems  are one  of the  most studied
materials  in physics owing  to their  complexity and  the competition
between various  ground states.  A  well-known example of this  is the
competition between charge density wave (CDW) and superconductivity in
layered transition  metal dichalcogenides (TMD) brought  back to focus
recently  by a  number of  experiments~\cite{aebi,dord} (see  Beck, et
al.,~\cite{beck_njp} for  a recent review).  The CDW  in these systems
was  thought to be  occuring via  Peierls~\cite{peier} mechanism  of a
divergence in the response function in systems with Fermi surface (FS)
nesting.  Despite experimental  and theoretical efforts, understanding
CDW and  related physical responses  of a two dimensional  layered TMD
remains  questionable  for  thirty  years  now.  Recent  discovery  of
superconductivity  on intercalation and  under pressure  has rekindled
the interest  in them.  Presence of unfilled  or partially  filled $d$
band and  an almost filled $p$  band near Fermi level,  with intra and
inter-band  interaction  on a  triangular  lattice, produce  competing
broken  symmetry  states. A  detailed  understanding  of these  broken
symmetry states  in quantum matter is,  therefore, at the  heart of an
understanding of the TMD as well~\cite{atprl}.

2H-NbSe$_2$ is  one of the  first 2D CDW  systems known~\cite{wilson},
becomes a superconductor at 7.2 K and was, in fact, considered a "high
$T_c$  superconductor"  prior to  the  cuprate-revolution. Even  after
about  forty  years, the  mechanism  of  CDW  in this  system  remains
shrouded  in mystery~\cite{dord,straub,shen,johannes08}.   In contrast
to another, similar, 2H-type  dichalcogenide, 2H-TaSe$_2$, which has a
continuous transition from a normal  to an incommensurate CDW state at
122  K and then  a first  order lock-in  transition to  a commensurate
phase at 90  K, NbSe$_2$ has a second order  phase transition from the
normal  to only  a single  CDW state  at T$_{CDW}$=33~K.  This ordered
state finally shows superconductivity below  7 K. The search for Fermi
surface (FS) nesting in the CDW transition of NbSe$_2$ has been a very
strenuous  one~\cite{straub}, never showing  any strong  indication of
nesting at  FS~\cite{boris_prl,shen} or saddle  bands with appreciable
density  of  states   (DOS)~\cite{tonjes,shen}.  Moreover  the  charge
susceptibility calculation shows no sharp peak at the CDW wave vector,
q$_{CDW}$ = ($2\pi/3, 2\pi/3$)~\cite{johannes08,boris_prl}.  Though it
has a broad  peak in its real part, there  is no discernible signature
of the peak in  the imaginary part~\cite{johannes08}.  In addition, it
is  found  from  angle  resolved  photoemission  spectroscopy  (ARPES)
studies     that     FS      is     gapped     only     in     certain
directions~\cite{boris_prl,shen}  and   in  the  temperature-dependent
charge excitation spectra, there is  no indication of the opening of a
gap, although the possibility of  a pseudogap has been noted in ARPES,
X-Ray and  STM studies recently~\cite{utpal}.   Consequently, there is
no  pronounced   effect  of  CDW   transition  in  the   transport  in
2H-NbSe$_2$~\cite{dord},   apart   from   a  peak   developing   below
T$_{CDW}$. As in 2H-TaSe$_2$, the in-plane resisitivity of 2H-NbSe$_2$
is  monotonic and approximately  linear in  $T$ over  a wide  range of
temperatures.  The c-axis  resistivity has almost similar T-dependence
as the in-plane one, though the c-axis resistivity is much larger. The
anisotropy, $\rho_c(T)/\rho_{ab}(T)$,  is around 30 between  300 K and
100 K  and decreases to  about 10 at  10 K.  $\rho(T)$  of 2H-NbSe$_2$
shows only a weak change of slope at the CDW transition at 33~K, below
which an increased metalicity has been observed.

The  optical   conductivity  shows  a  Drude  peak   and  the  typical
`shoulder-like' mid-infrared  feature of a correlated  system at about
300~K.  Several  differences   between  the  two  2H  dichalcogenides,
2H-NbSe$_2$  and 2H-TaSe$_2$, can  be noticed  in the  scattering rate
data also.  The temperature dependence of the in-plane scattering rate
in 2H-NbSe$_2$ is weak.  Quite remarkably, the peak in the self-energy
in  the ARPES  excitation spectra  is located  at different  places at
different   energies,  indicating  a   weak  momentum   dependence  of
self-energy in  2H-NbSe$_2$, coming, presumably,  from electron-phonon
coupling~\cite{valla1}. Contrast  this to 2H-TaSe$_2$,  where the peak
in  the self  energy is  at a  range high  enough for  electron phonon
coupling.      Moreover,     neutron~\cite{moncton}      and     x-ray
scattering~\cite{murphy} phonon dispersion data show a softening close
to  the  CDW wave  vector  in  2H-NbSe$_2$,  lending credence  to  the
coupling  of  phonons  in  the  CDW transition  of  2H-NbSe$_2$.   The
electron-phonon interaction  being relevant  here, may also  cause the
superconductivity to appear at a `higher' temperature of around 7 K in
this  system, while  supecronductivity  is  found only  at  200 mK  in
2H-TaSe$_2$,      where       electron-phonon      interaction      is
insignificant~\cite{valla1,atprl}.  Additionally,  CDW transition does
not affect the optical conductivity either, the zero energy Drude-like
peak  and  its narrowing  with  decreasing  temperature  is a  generic
feature of 2D dichalcogenides.

All this indicates the formation of a pseudogap at a temperature above
$T_{CDW}$  as  revealed  in  the  recent  experiments~\cite{utpal}  in
2H-NbSe$_2$   and  theoretical   analysis  of   TMD~\cite{atprl}.  The
progressive narrowing of the Drude  peak in the far infrared region on
lowering  the  temperature   coexists  with  mid-infrared  absorption.
Considering  all, these  results point  to  the inadequacy  of the  FS
nesting  scenario  based  on  LDA  studies.  If  FS  nesting  were  in
operation, then FS and transport data  would show opening of a CDW gap
at the reconstructed FS and  a severe drop in conductivity entering in
to the  commensurate CDW state,  which are in conflict  with available
experimental results.  Besides, the band picture  cannot be reconciled
with the  temperature dependent  large spectral weight  transfer (SWT)
observed. A theory  should fulfil all these criteria  and describe the
smooth  crossover of  the normal  state to  a CDW  phase.  A preformed
excitonic  liquid  (PEL)  scenario ~\cite{atprl},  which  successfully
described   2H-TaSe$_2$  and   1T-TiSe$_2$  (with   the   addition  of
electron-lattice coupling in  the latter) can resolve this  if the PEL
appears at high $T$. This can condense into a CDW state at low $T$ via
electron-lattice coupling.

\section{Method}
We propose  a preformed excitonic liquid scenario  for 2H-NbSe$_2$ and
show that a wide range of normal state properties could be understrood
within this picture. We  use exciton-phonon coupling for the emergence
of  an unconventional  CDW state  at lower  T as  instabilities  of an
incoherent preformed excitonic liquid via exciton-lattice interaction.
\begin{figure}[!ht]
\centering
{\includegraphics[angle=0,width=\columnwidth]{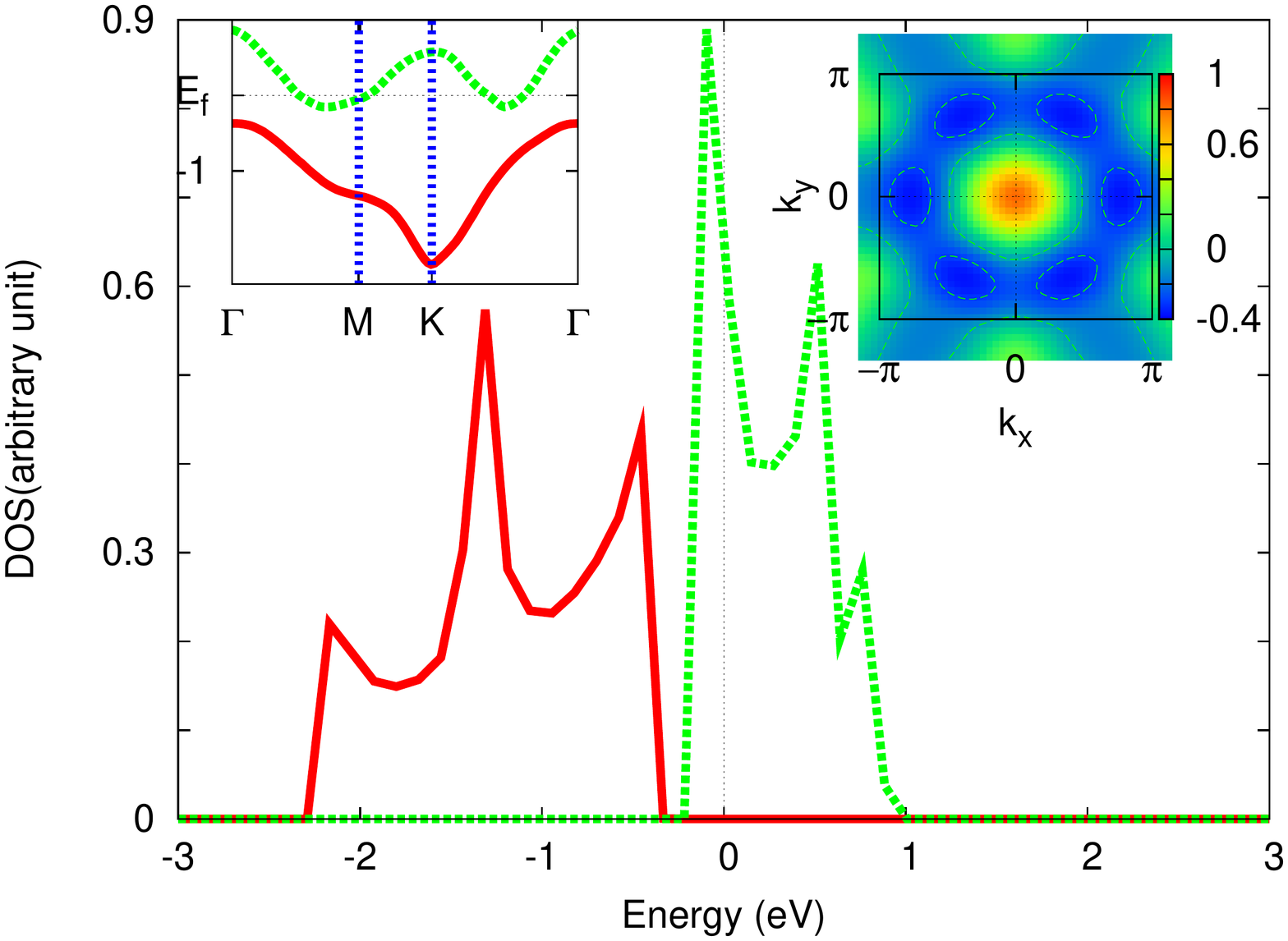}}
\caption{(Color Online) Non interacting DOS (main panel), bands closest to FL (left
inset) and Fermi surface (right inset) calculated from a tight binding fit (see text). 
The red band represents the band with predominantly Se-p character and the  
green band is predominantly of Nb-d character.}
\label{fig1}
\end{figure}
\begin{figure}[!ht]
\centering
(a)
{\includegraphics[angle=0,width=0.8\columnwidth]{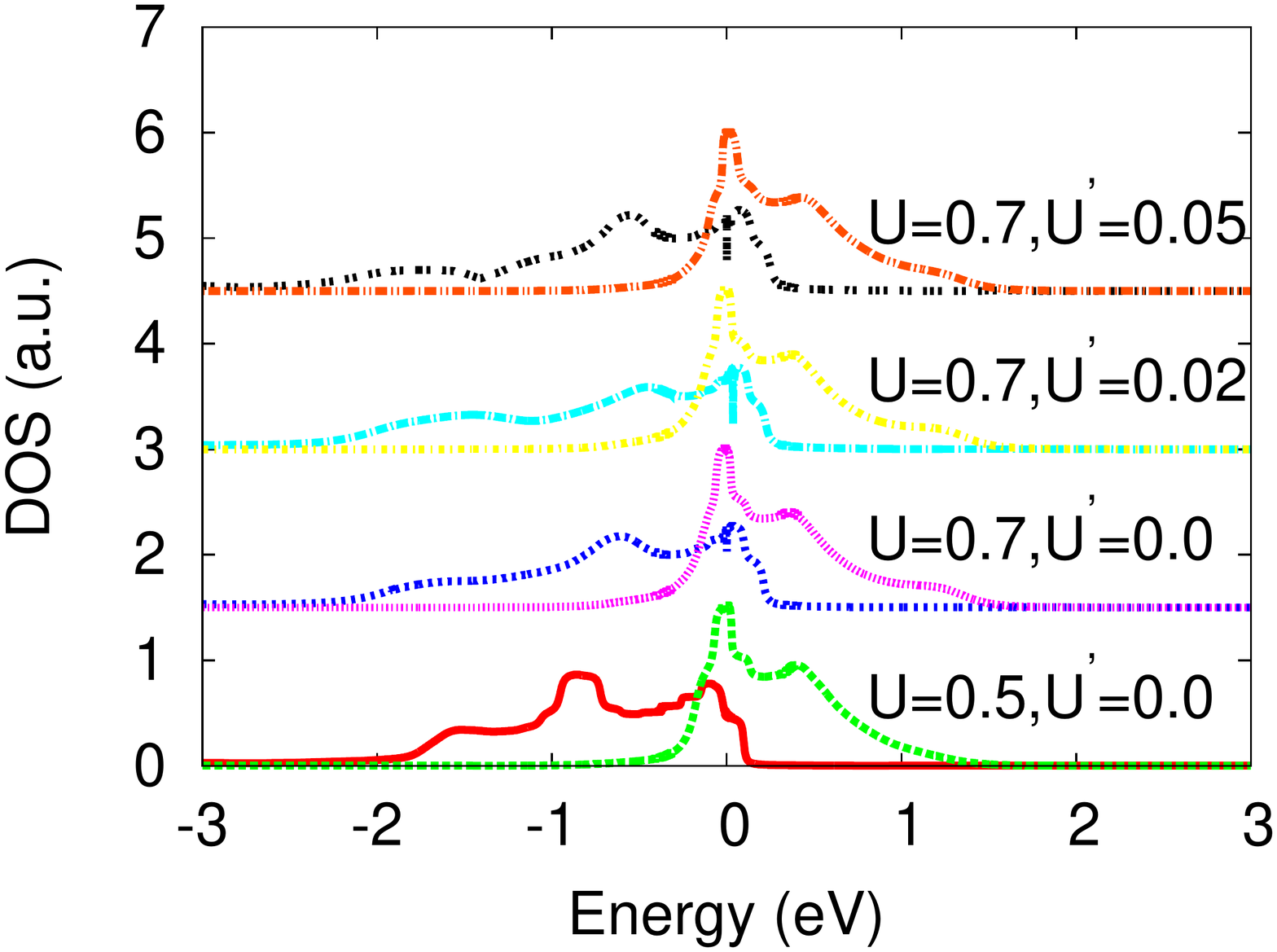}}

(b)
{\includegraphics[angle=0,width=0.8\columnwidth]{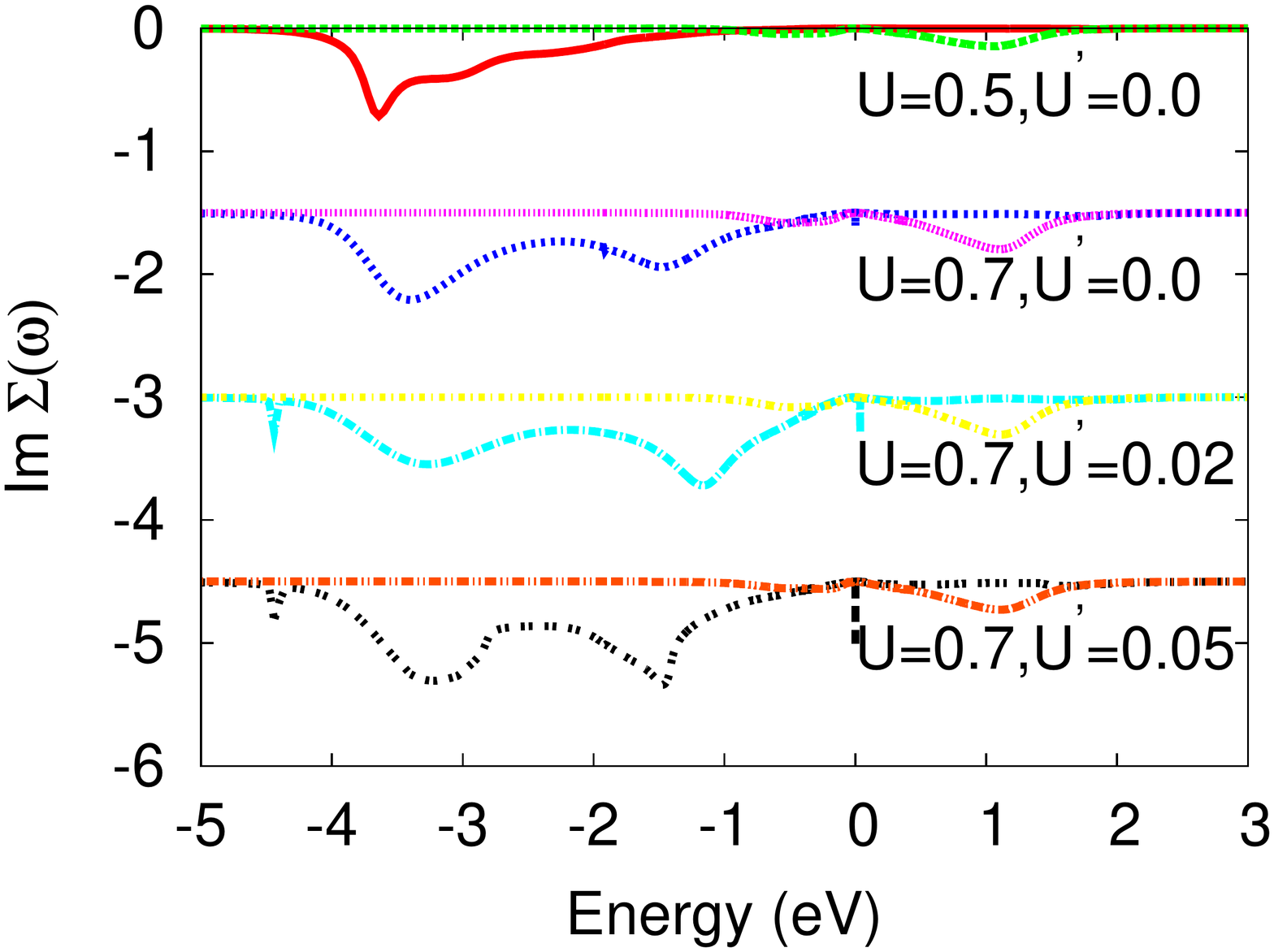}}

(c)
{\includegraphics[angle=0,width=0.8\columnwidth]{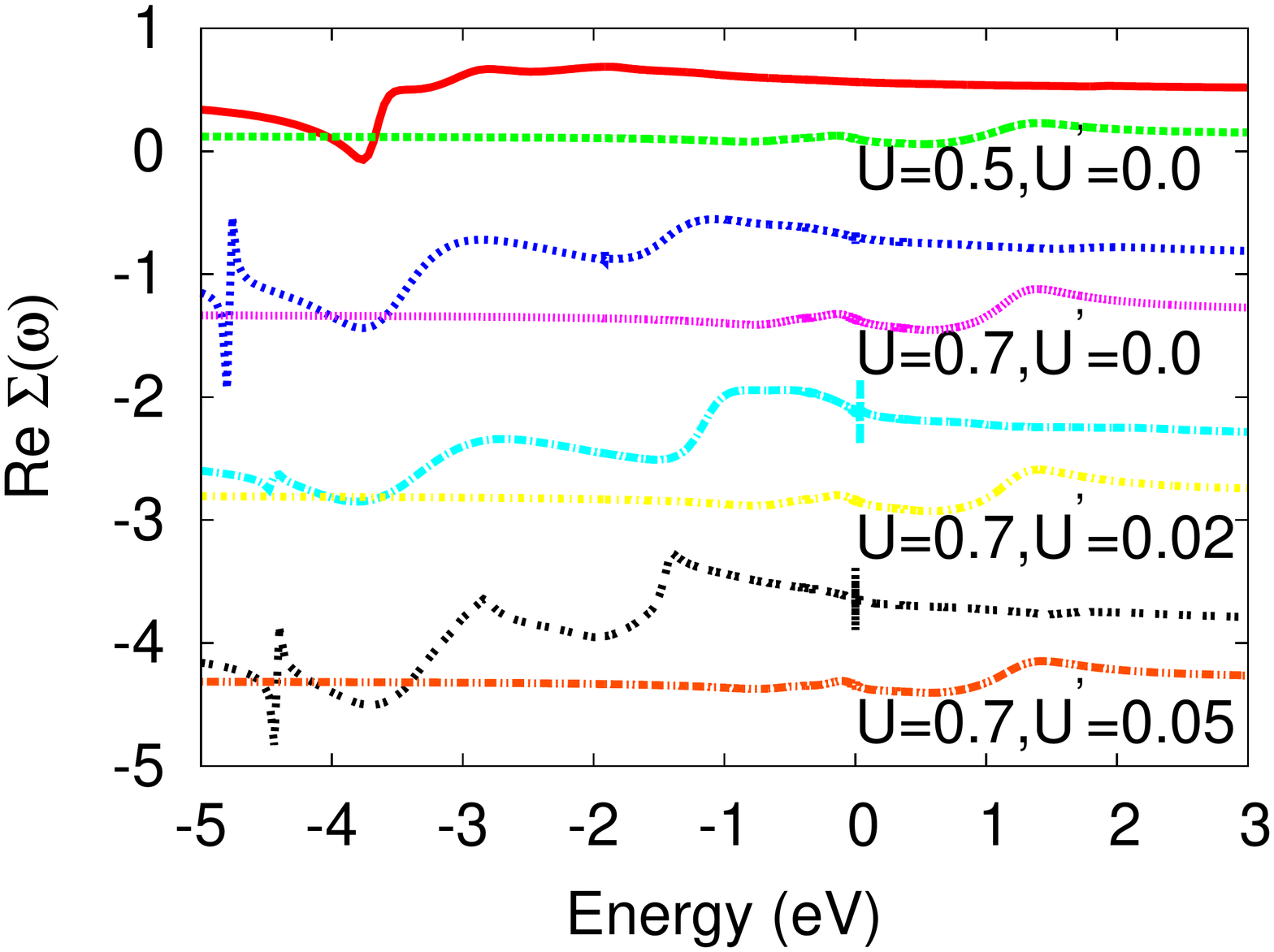}}
\caption{(Color Online)  DMFT (a) DOS  of Nb-$d$ band and  Se-$p$ band
  for several U and U$^{'}$ (=U$_{ab}$).  The DOS with peak near FL is
  from Nb-$d$ band. (b) Imaginary and (c) real part of self energy for
  the same values of U and U$^{'}$ as in (a). The self energies with a
  dip-like structure close to -1 eV  in the imaginary part are the $p$
  band self-energies.}
\label{fig2}
\end{figure}
\begin{figure}[!ht]
\centering
(a)
{\includegraphics[angle=0,width=0.8\columnwidth]{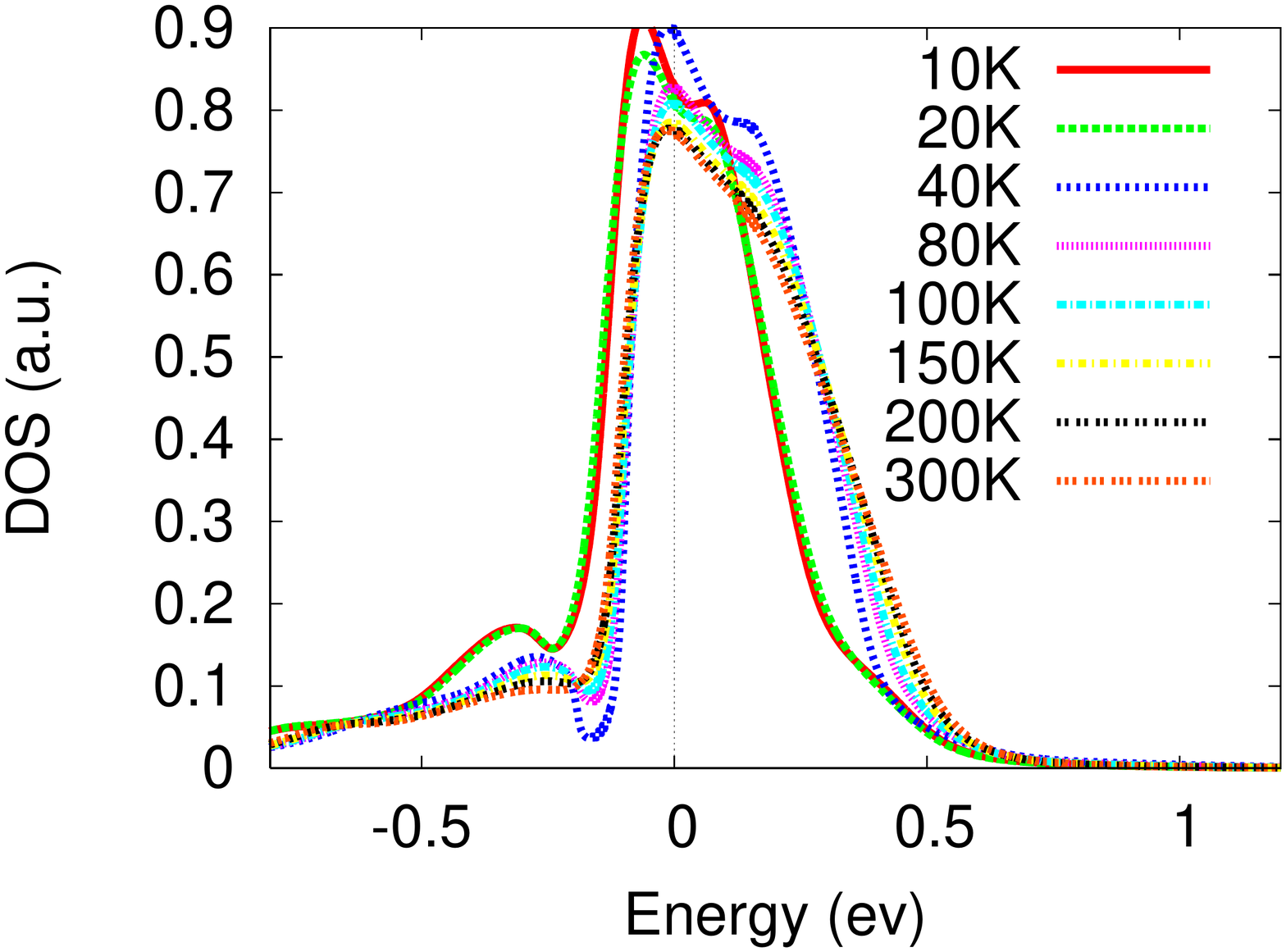}}

(b)
{\includegraphics[angle=0,width=0.8\columnwidth]{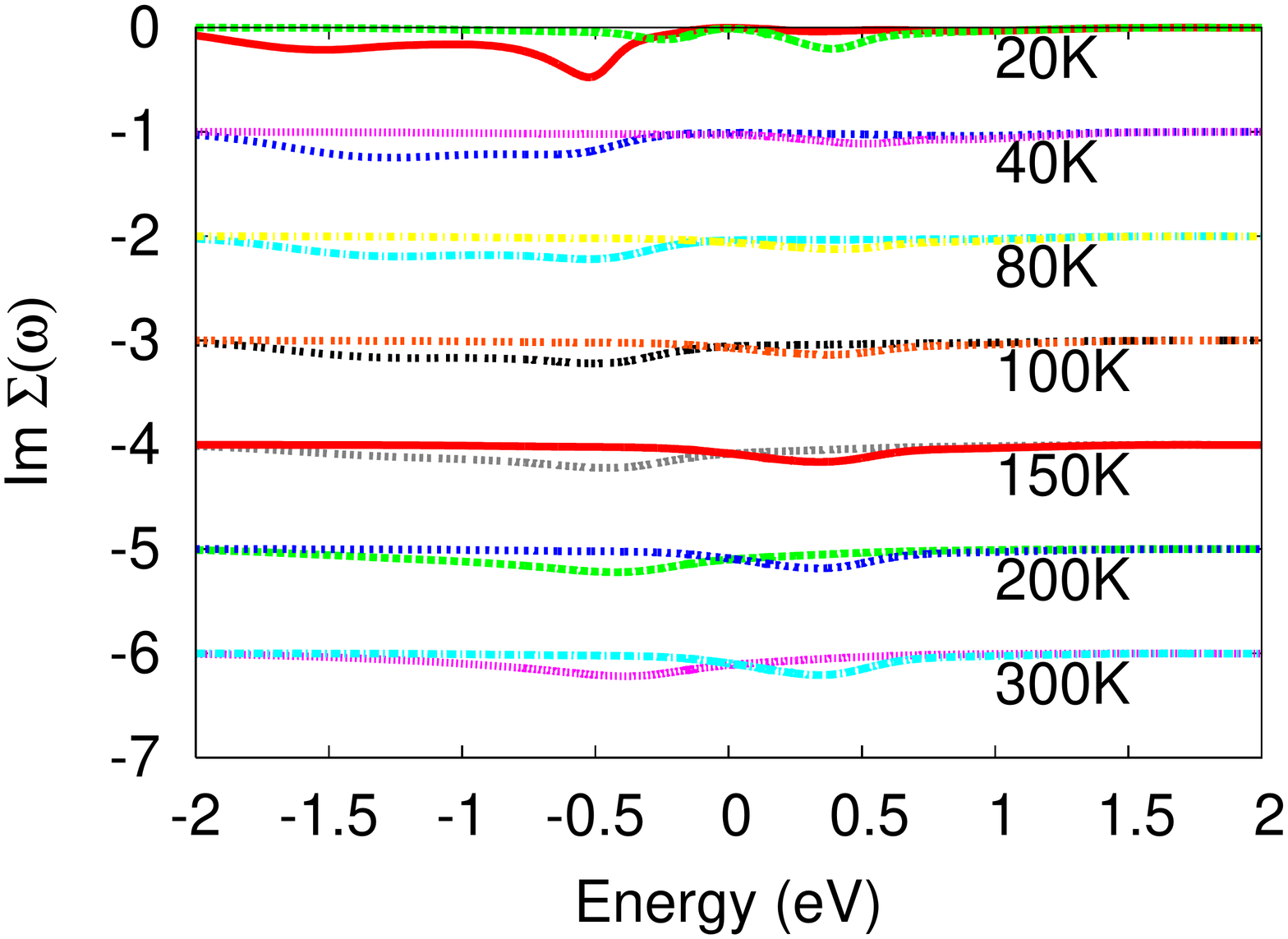}}

(c)
{\includegraphics[angle=0,width=0.8\columnwidth]{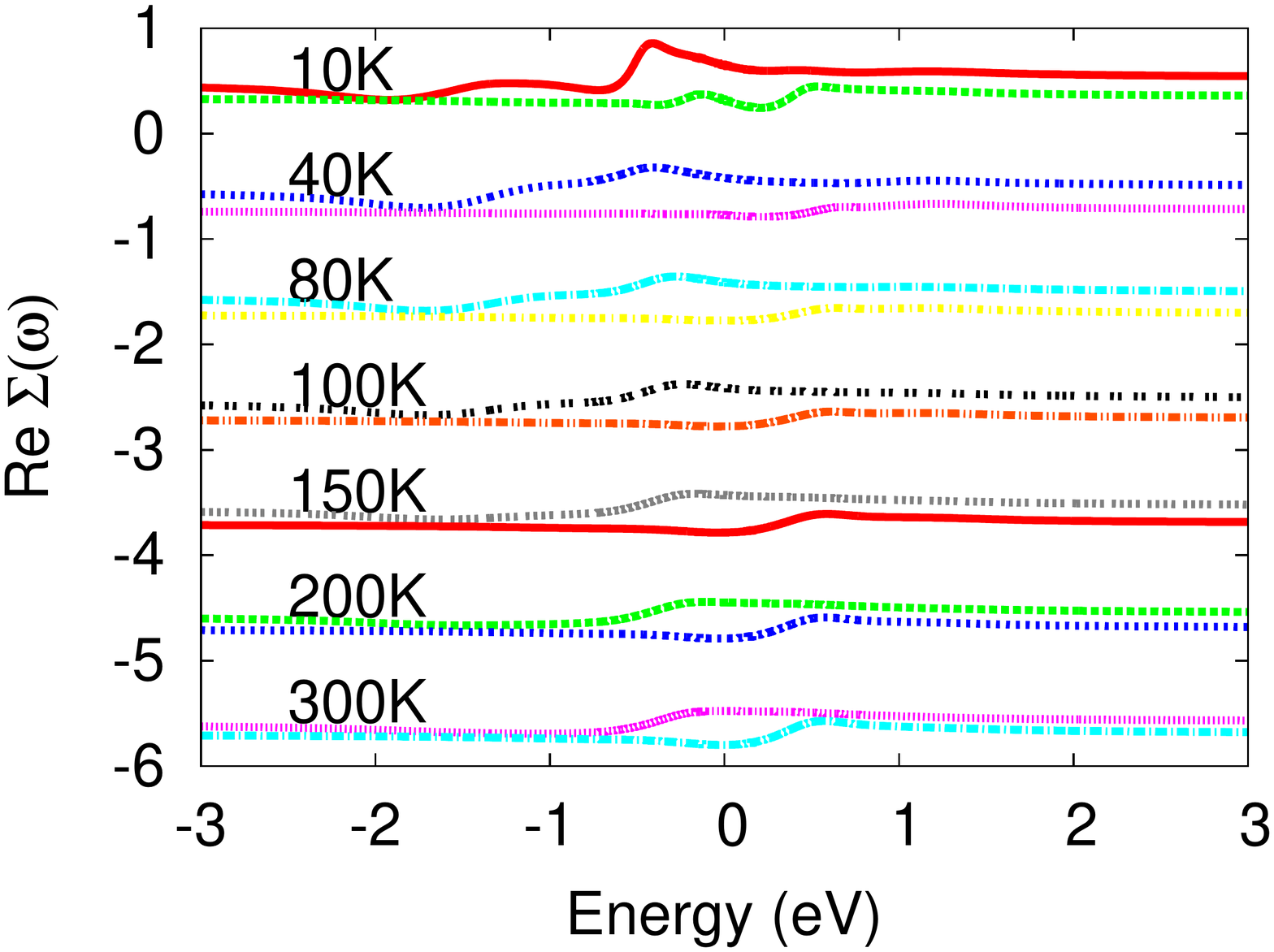}}
\caption{(Color  Online)  Temperature  evolution  with  U  =  0.5  eV,
  U$_{ab}$ = 0.05  eV, $t_{ab}=$0.1 eV and phonon  coupling 0.05 eV of
  (a)  DOS  of  Nb-$d$  band.  (b)  Imaginary and  (c)  real  part  of
  self-energy  of p and  d band.   The self-energies  with a  dip like
  structure  close to -0.5  eV in  the imaginary  part are  for Se-$p$
  band.}
\label{fig3}
\end{figure}
\begin{figure}[!ht]
\centering
\includegraphics[angle=0,width=0.8\columnwidth]{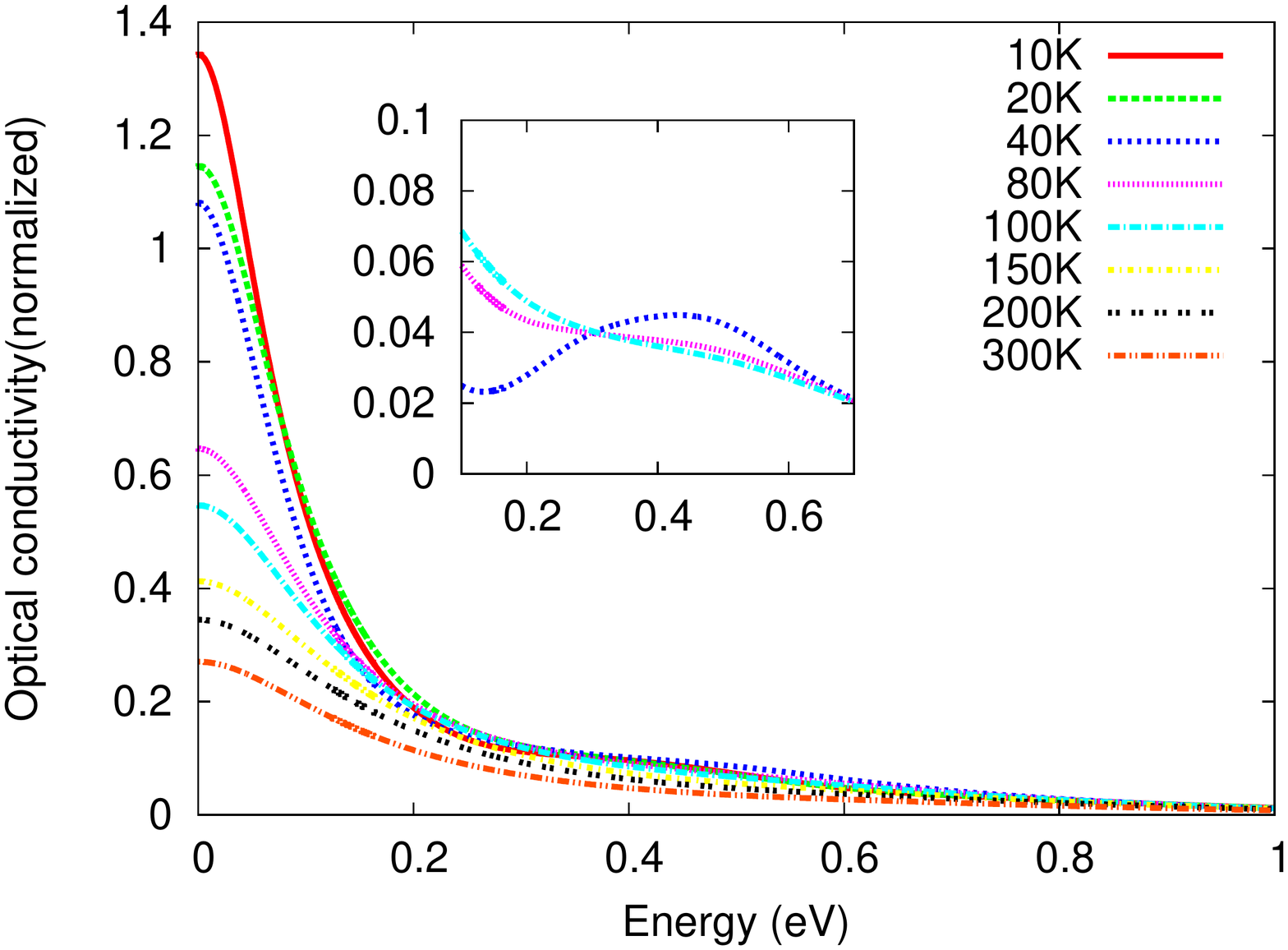}
\caption{(Color  Online) Optical  Conductivity at  various temperature
  from DMFT.  The same (inset) at low energy showing extended shoulder
  like feature in an enlarged scale.}
\label{fig4}
\end{figure}
\begin{figure}[!ht]
\centering
\includegraphics[angle=0,width=0.8\columnwidth]{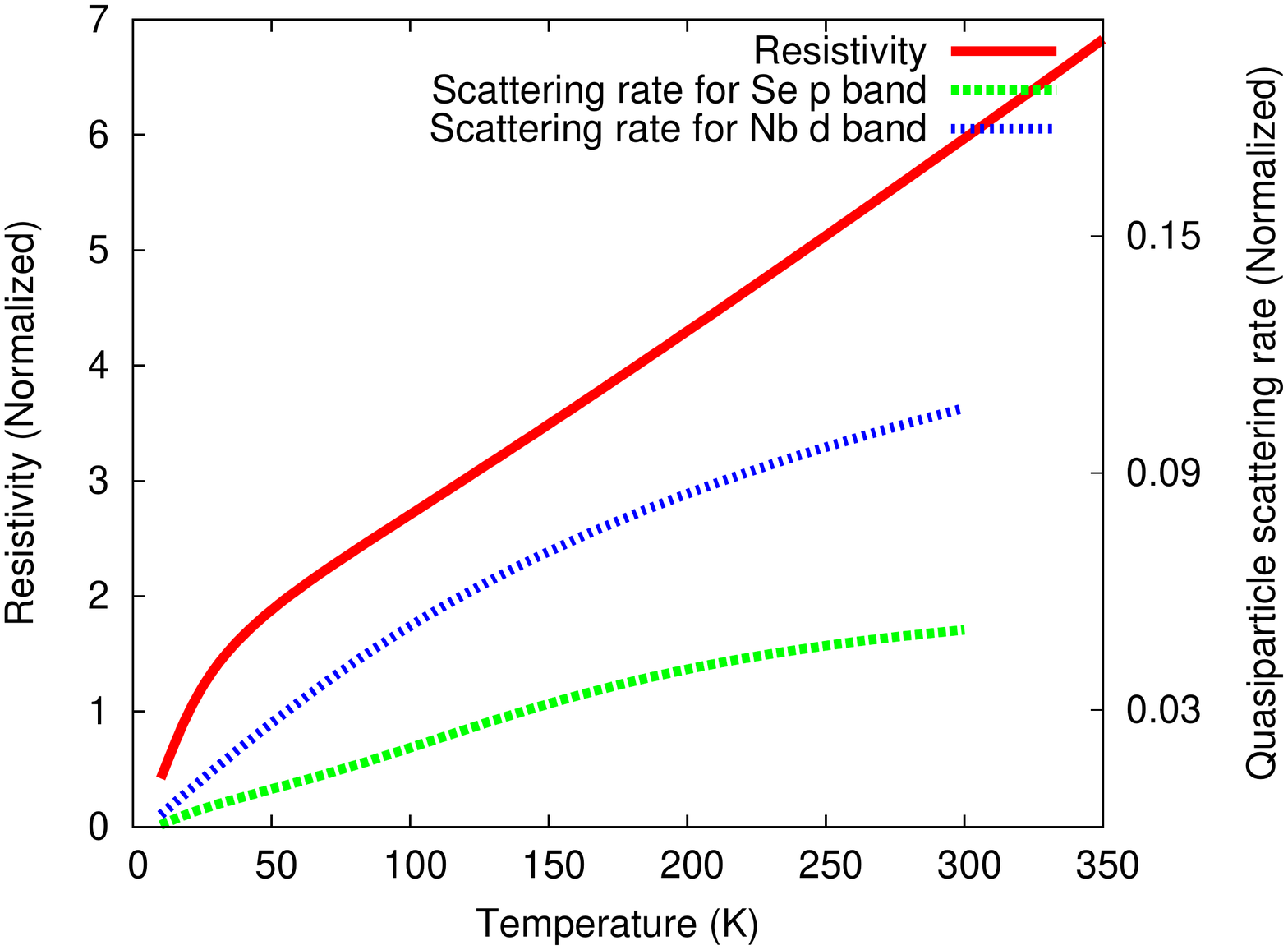}
\caption{(Color  Online) Temperature  dependent  dc resistivity  (left
  side scale) and scattering rates for the two bands (the scale on the
  right side).}
\label{fig5}
\end{figure}
\begin{figure}[!ht]
{\includegraphics[angle=360,width=0.8\columnwidth]{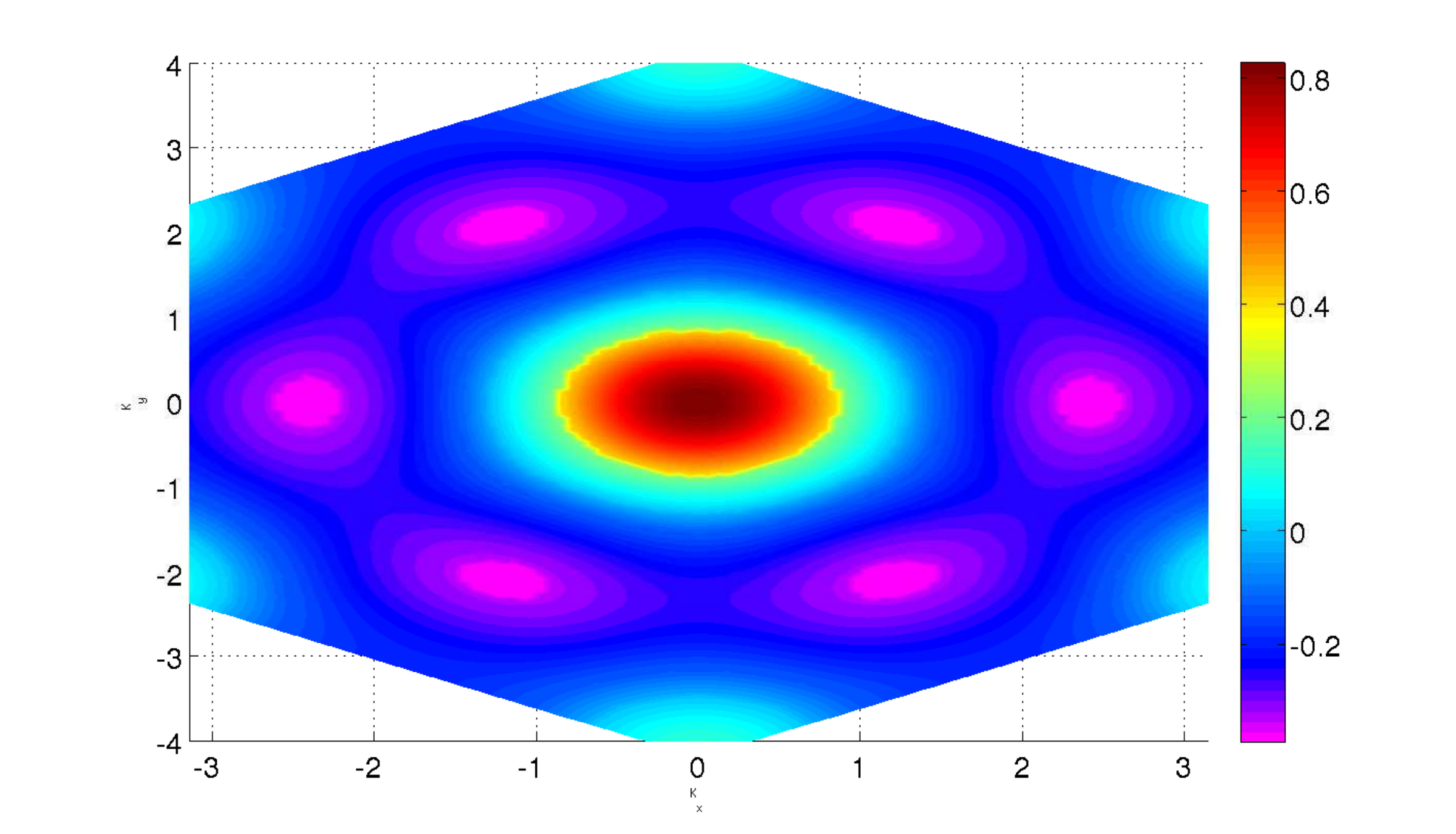}}
\caption{(Color Online) The interacting Fermi surface at 20K.}
\label{fig6}
\end{figure}
\begin{figure}[!ht]
(a)
\includegraphics[angle=0,width=0.8\columnwidth]{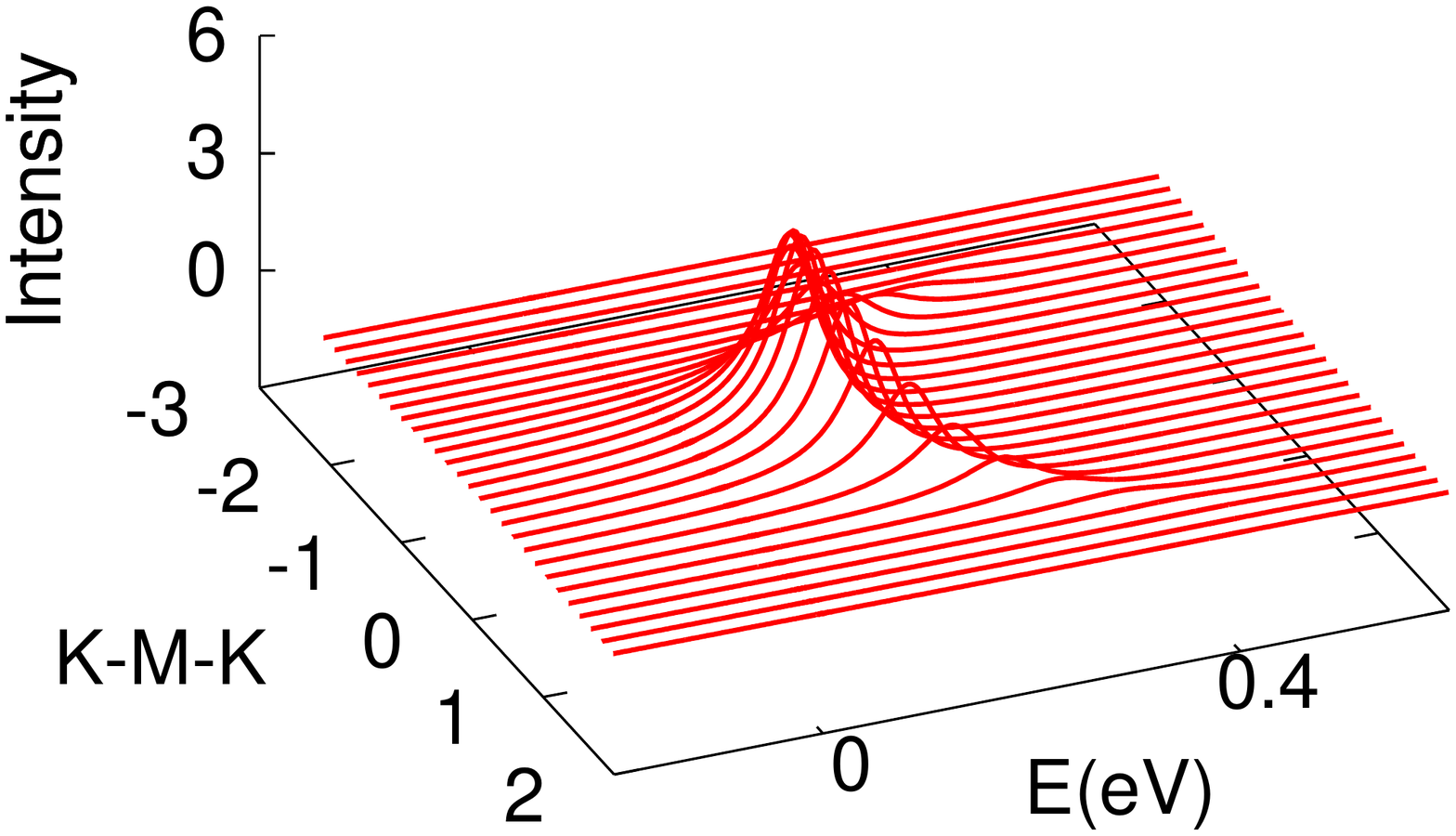}

(b)
\includegraphics[angle=0,width=0.8\columnwidth]{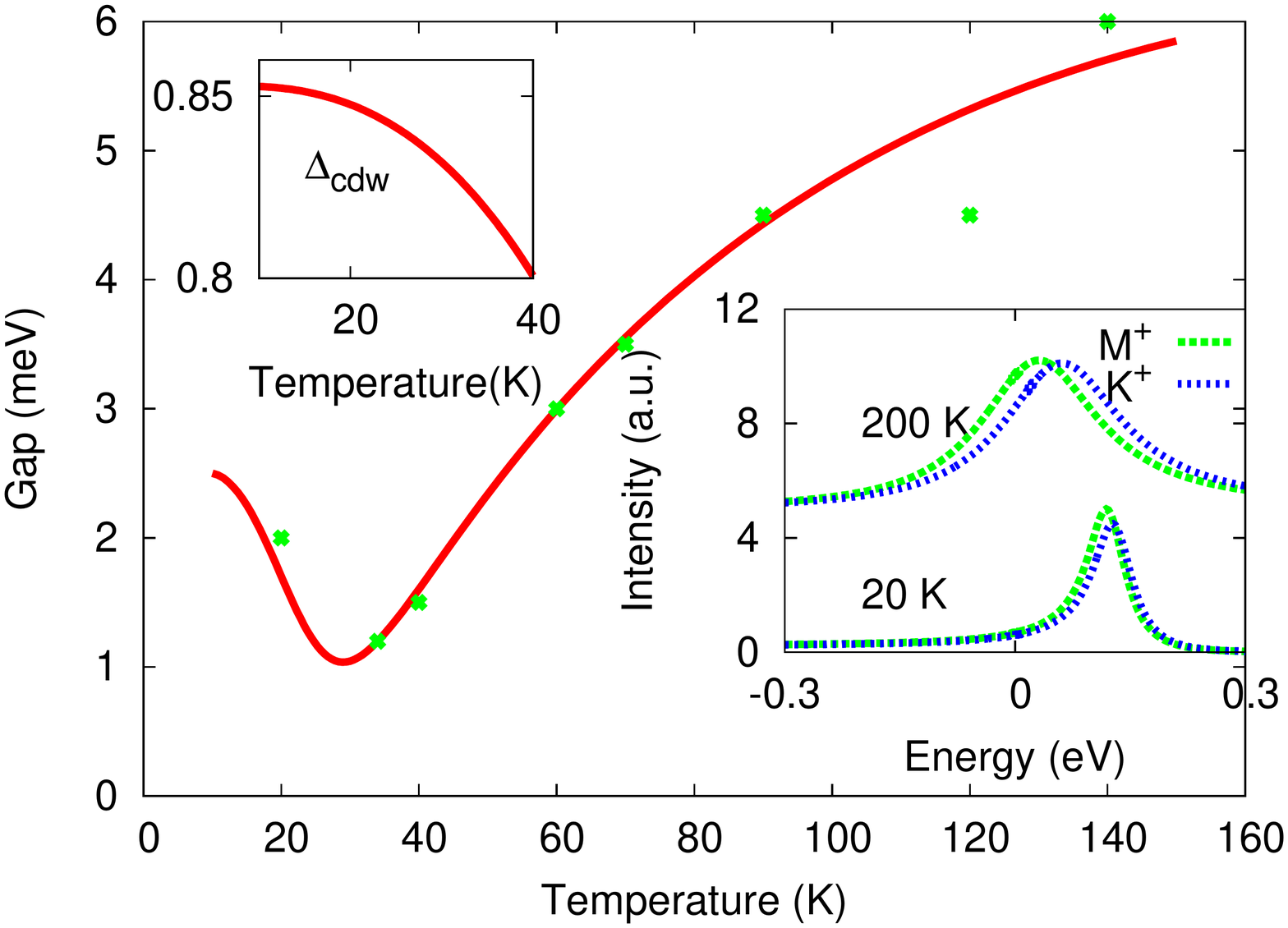}
\caption{(Color   Online)    (a)   Energy-momentum   distribution   of
  intensities along  K-M-K high  symmetry direction. (b)  Variation of
  Gap   (as  defined  by   Borisenko  et   al.~\cite{boris_prl})  with
  temperature.    The     green    dots    are     the    experimental
  points~\cite{boris_prl} for  comparison. Inset of  lower panel shows
  energy distribution curves  (EDC) at 20 K and 200  K along K$^+$ and
  M$^+$ direction (again,  as defined in Borisenko, et  al.) The inset
  on top shows CDW amplitude with temperature.}
\label{fig7}
\end{figure}
\begin{figure}[!ht]
\centering
\includegraphics[angle=0,width=0.8\columnwidth]{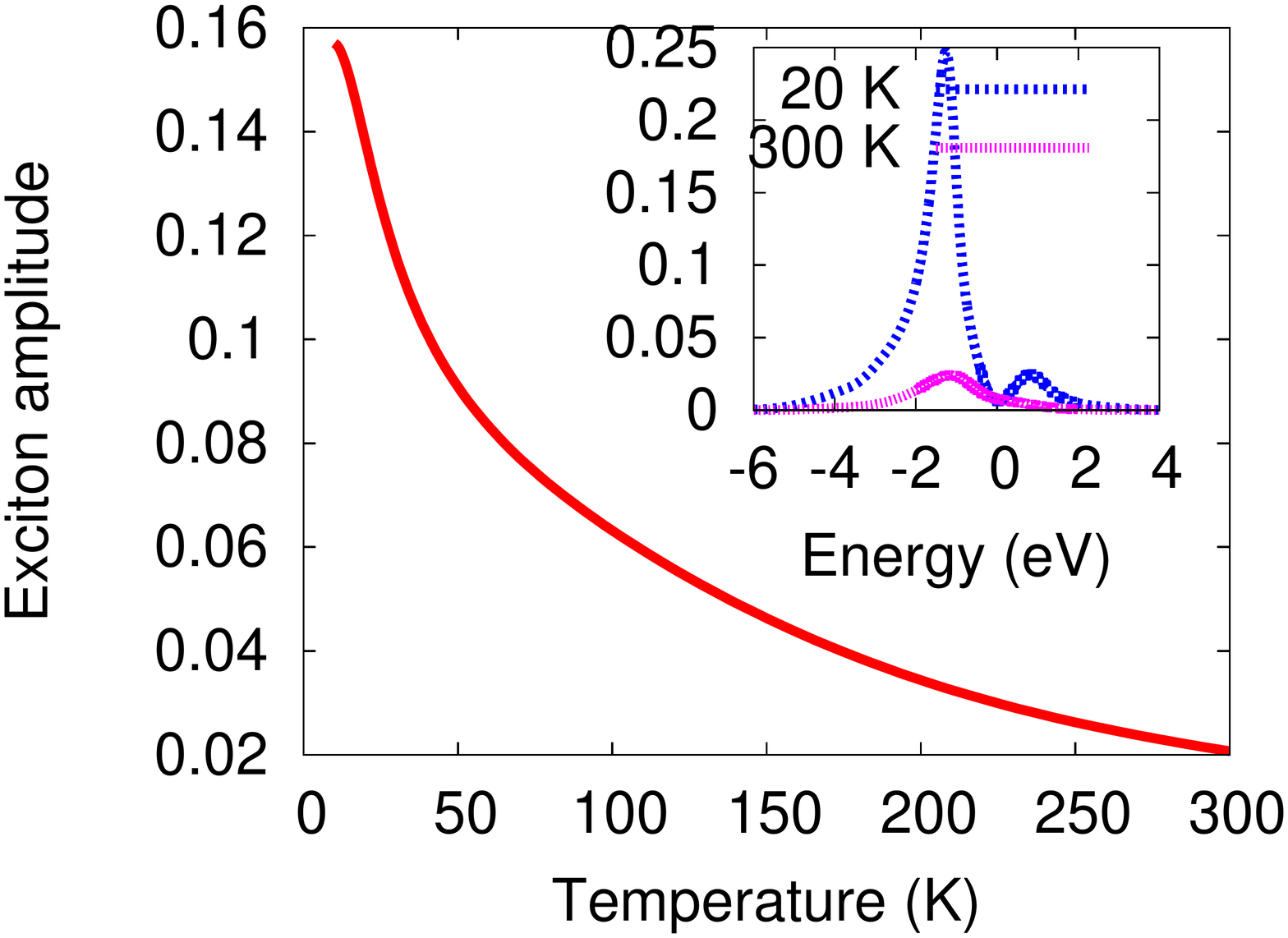}
\caption{(Color Online) Inter-orbital excitonic order parameter. Inset
  shows imaginary  part of the  off-diagonal Green's function at  20 K
  and 300 K.}
\label{fig8}
\end{figure}

\noindent  The  tight  binding  band structure  of  2H-NbSe$_{2}$  was
constructed  from  a  22$\times22$  matrix  using  Nb-$d$  and  Se-$p$
states. We show  the two bands closest to  $E_{F}$ (with predominantly
Nb-$d_{z^2}$  and   Se-$p_{z}$  character)   in  top  left   inset  of
Fig.~\ref{fig1} ,  as well  as the  FS (top right  inset) and  the DOS
(main panel), in fairly good agreement with LAPW results~\cite{doran}.
Only  one, nearly  spherical,  FS sheet  can  be seen  in  the LDA  FS
(Fig.~\ref{fig1} right  inset) which is already  in contradiction with
the notion  of FS nesting for  CDW formation.  In the  TMD systems the
conduction  d-band is nominally  either half-filled  or empty  and the
valance  band is  nominally  filled. The  LDA  shows partially  filled
conduction  and  valence  bands  with  low carrier  density.  In  this
situation,  the  formation  of  exciton  comes  about  via  a  sizable
$d_{z^2}-p_z$ mixing, which hybridises  the electrons and holes in the
$d$ and  $p$ bands in  conjunction with dynamical  renormalization and
electron-hole attraction by U$_{ab}$. Besides, at finite temperatures,
there are  thermally generated  holes in the  valence band  aiding the
excitonic instability.  In this situation a small electron correlation
will  form excitons  at high  $T$.  The  electronically  active states
comprise Nb-$d_{z^2}$  and Se-$p_{z}$  bands and $U\simeq  0.7$~eV and
$U_{ab}\simeq0.05$~eV  are the  typical values  of  interactions where
exciton  formation is  found to  be most  likely. In  view of  the bad
metallic normal  state, the ordering instabilities of  such bad metals
{\it  cannot},  by  construction,  be  rationalised  by  appealing  to
static-mean-field  theory:  this  can  only be  reliably  accessed  by
approaches  which  capture  dynamical  correlations  adequately.   For
2H-NbSe$_{2}$, a minimal two-band correlated model as defined below is
mandated by tight binding results, and adequate treatment of dynamical
correlations   underlying   incoherent   behavior   is   achieved   by
dynamical-mean field theory (DMFT).  The  DMFT is now proved to be one
of the  most successfull approaches to  correlated electronic systems,
making  them attrative  tools  to  use in  the  present context.   The
two-band  Hubbard model  we  use is  ($d_{z^2}=a$  and $p_{z}=b$)  
\be
\begin{split}
H_{el}&=\sum_{{\bf k},l,m,\sigma}(t_{\bf k}^{l\ne m}+\epsilon_{l}\delta_{lm})c_{{\bf     k}l\sigma}^{\dag}c_{     {\bf
    k}m\sigma}+U\sum_{i,\mu}n_{i\mu\uparrow}n_{i\mu\downarrow}\\
&+ U_{ab}\sum_{i}n_{ia}n_{ib}
\end{split}\vspace{-2em}
\ee
\noindent  where  $l,\,m,   $  take  values  a,  b   and  $U$  is  the
intra-orbital Coulomb  correlation (taken to  be same for  $a,b$ being
small compared to band-width (see later); we have checked that results
are insensitive to this  choice within reasonable limits); $U_{ab}$ is
the  inter-orbital correlation.  Further,  in TMD,  the most  relevant
$A_{1g}$ phonon  mode couples to the inter-band  excitons by symmetry,
and      the      electron-phonon      coupling     is      $H_{el-l}=
g\sum_{i}(A_{i}+A_{i}^{\dag})(c_{ia}^{\dag}c_{ib}+h.c)$.    To   solve
$H=H_{el}+H_{el-l}$  within DMFT, we  have combined  the multi-orbital
iterated perturbation theory (MOIPT) for $H_{el}$~\cite{laad} with the
DMFT for  polarons by Ciuchi  {\it et al.}~\cite{ciuchi}.  Finally, we
address the broken  symmetry phase quite differently~\cite{atprl} from
the LDA  based pictures involving  band folding.  This is  called for,
from  the above  discussions and  LCAO+DMFT results  below,  since the
`normal' state  is found to be  an incoherent PEL  at high temperature
without well  defined Landau quasiparticles. Instability  to CDW order
then  cannot occur  via the  traditional band-folding  involving Fermi
liquid  quasiparticles.  Rather,  since  {\it  coherent}  one-particle
inter-band  mixing  is  inoperative,  ordered states  must  now  arise
directly as  two-particle instabilities  of the bad  metal. This  is a
situation  reminiscent  of  2H-TaSe$_2$  \cite{atprl}  and  we  employ
similar arguments here.  So for 2H-NbSe$_2$ as well,  the two particle
interaction is  more relevant than the  incoherent one-electron mixing
and  therefore   we  construct  an   effective  Hamiltonian  involving
two-particle processes:
\newline                                                 $H_{res}\simeq
-t_{ab}^{2}\chi_{ab}(0,0)\sum_{<i,j>,\sigma\sigma'}c_{ia\sigma}^{\dag}
c_{jb\sigma}c_{jb\sigma'}^{\dag}c_{ia\sigma'}$,  with  $\chi_{ab}$ the
inter-orbital susceptibility. Decoupling this intersite interaction in
a  generalised  HF sense  we  will  get  two competing  instabilities:
$H_{res}^{(HF)}=-\sum_{<i,j>,\sigma\sigma'}(\Delta_{1b}c_{ia\sigma}^{\dag}c_{ia\sigma}
+  \Delta_{ab}c_{ia\sigma}^{\dag}c_{jb-\sigma}^{\dag}  +  a\rightarrow
b)$,    with   $\Delta_{cdw}=(\Delta_{1a}-\Delta_{1b})\simeq   \langle
n_{a}-n_{b}\rangle$ representing a  CDW and $\Delta_{ab}\simeq \langle
c_{ia\sigma}c_{jb-\sigma}\rangle$     a     multiband     spin-singlet
superconducting  (SC)  order  parameters.   In what  follows,  at  low
temperature, we only concentrate on the broken symmetry CDW phase. The
singlet SC instability is an issue we will take up in the future. What
is  clear is  that, owing  to the  characteristic form  factor  of the
coefficient in the two-fermion  term in the effective Hamiltonian, the
SC  order  parameter  is  expected  to  be  anisotropic.  In  the  low
temperature phase,  we ignore  this unconventional SC  order parameter
and use the mean-field decoupling  of the effective Hamiltonian in the
particle-hole channel and calculate  the order parameter and transport
in the CDW phase at low $T$.

\section{Results} 
Going  back   to  Hamiltonian   Eq.(1)  and  following   the  standard
multi-orbital    DMFT    procedure    with    iterated    perturbation
scheme~\cite{note} as the impurity solver~\cite{atprl}, we compute the
DMFT spectral  function and transport properties in  the normal state.
We first work out the case without electron-phonon coupling.
\begin{figure*}
\centering
(a)
{\includegraphics[angle=0,width=0.5\columnwidth]{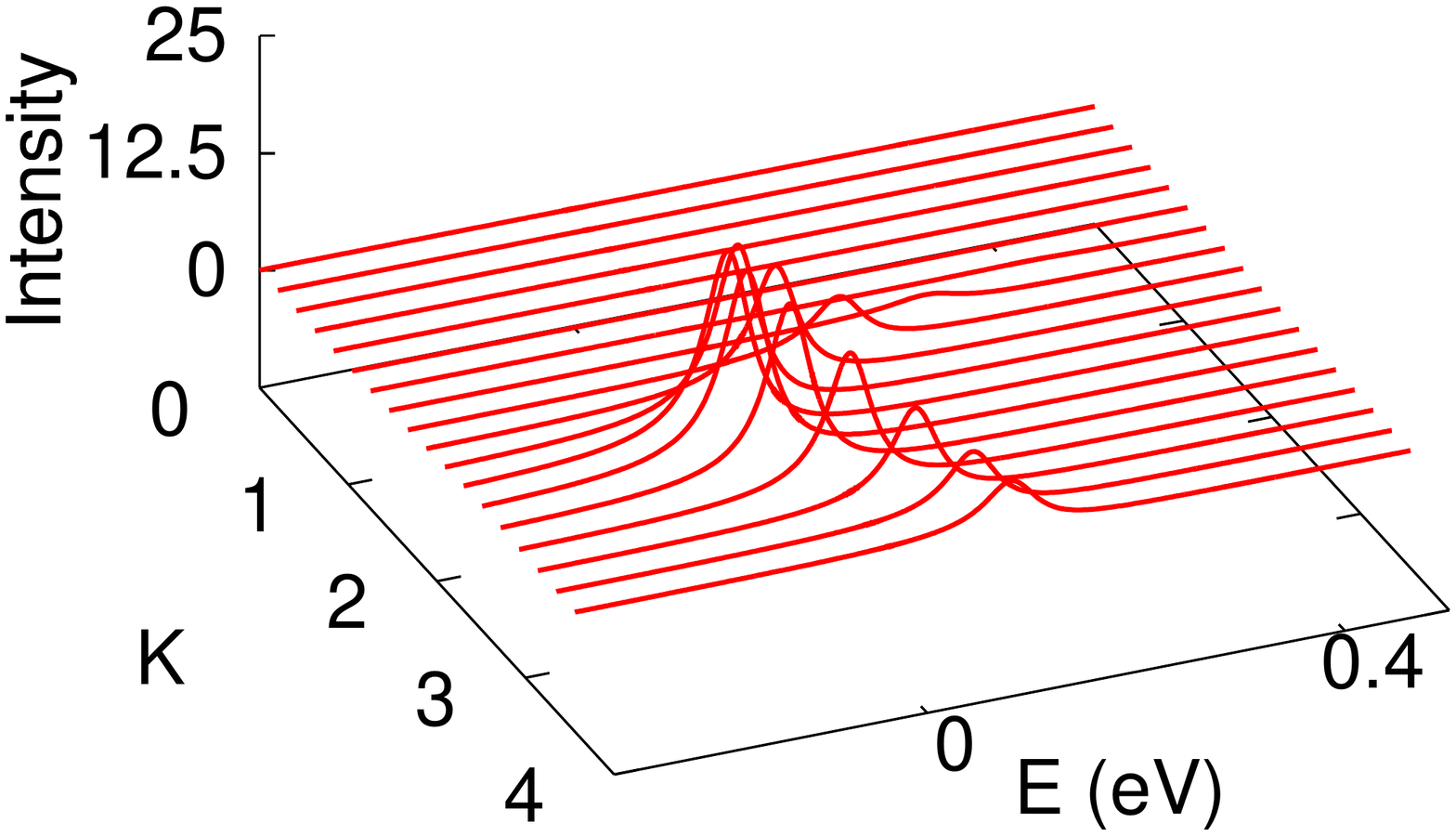}
}
(b)
{\includegraphics[angle=0,width=0.5\columnwidth]{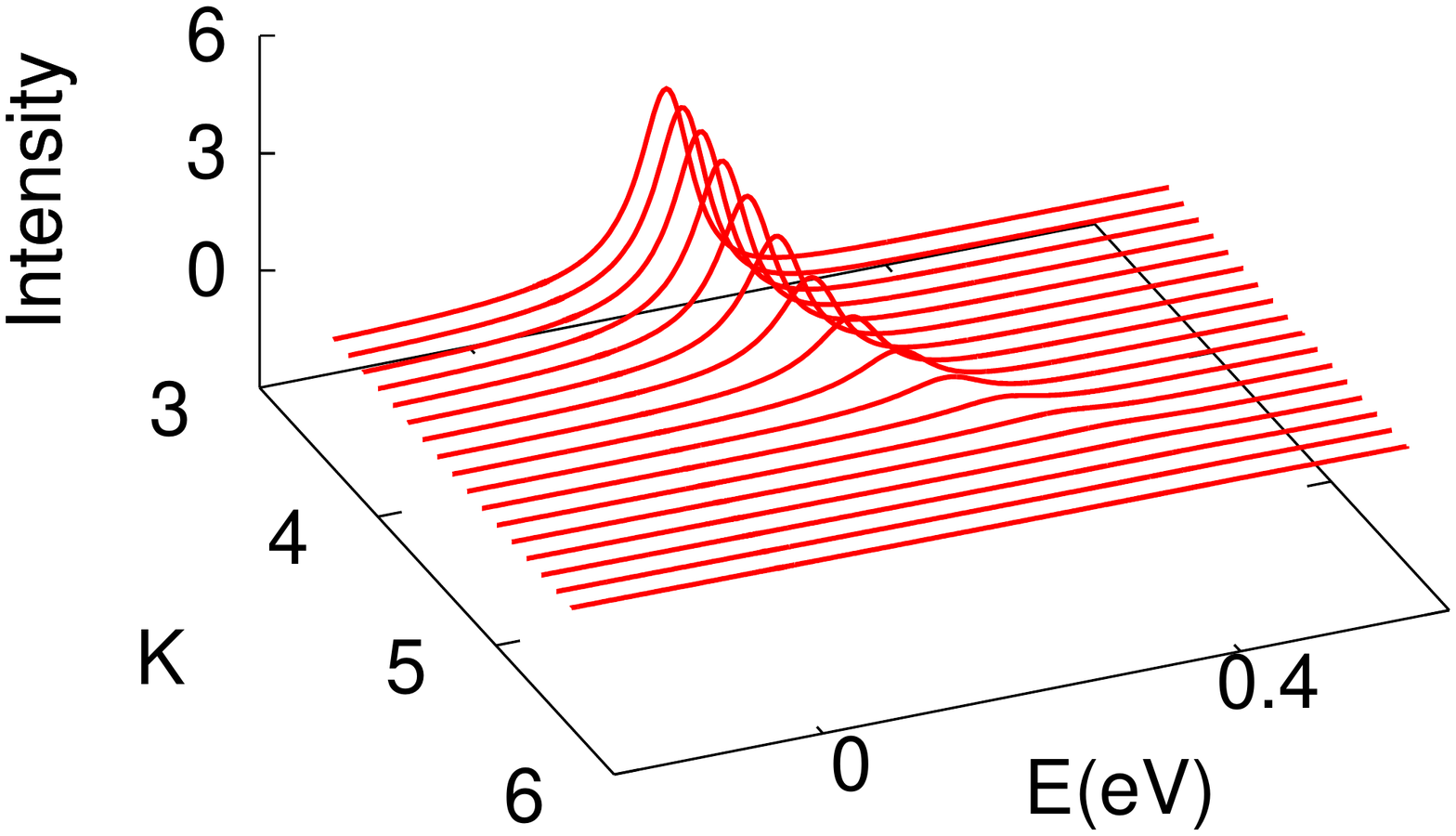}
}
(c)
{\includegraphics[angle=0,width=0.5\columnwidth]{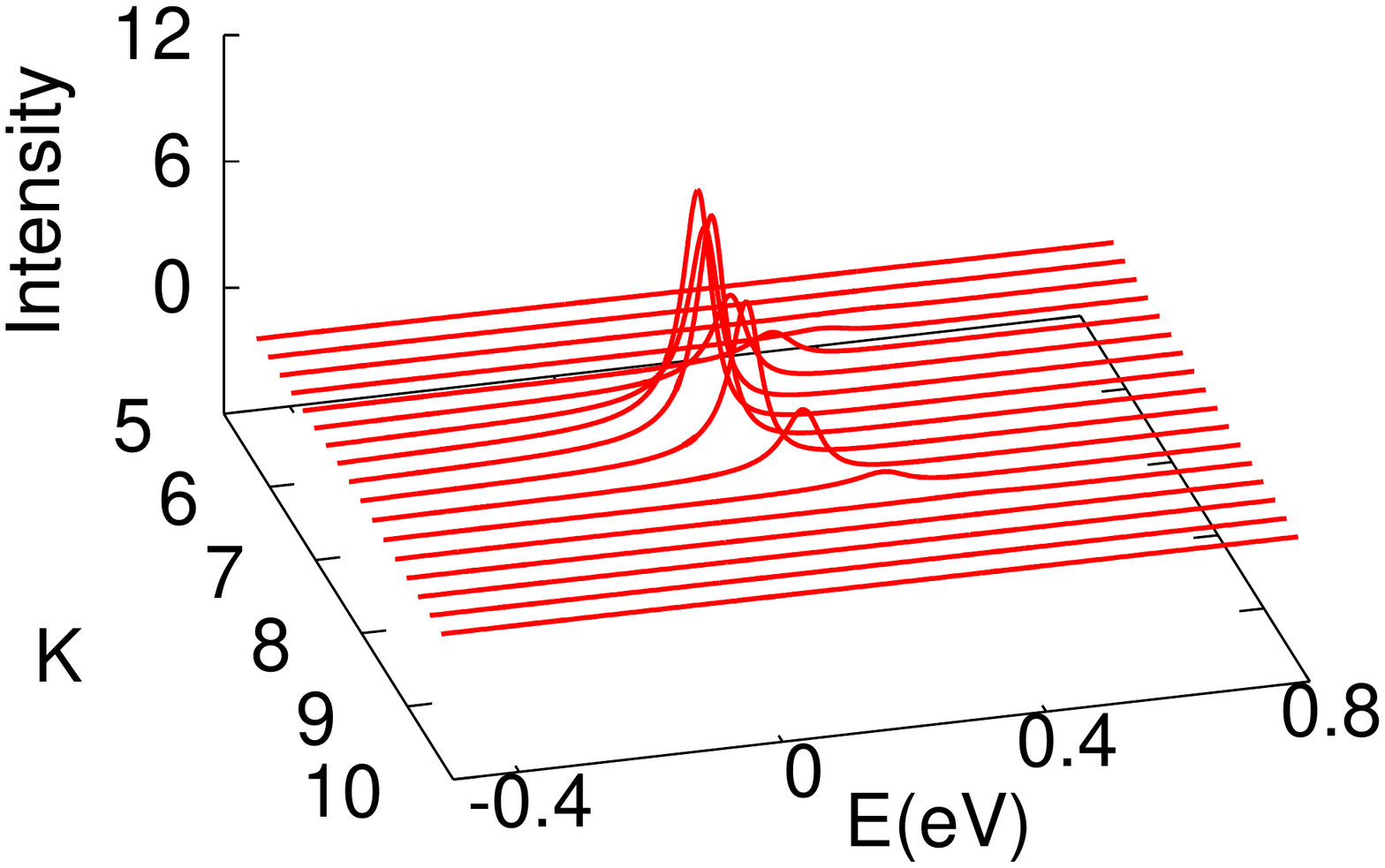}}
(d)
{\includegraphics[angle=0,width=0.5\columnwidth]{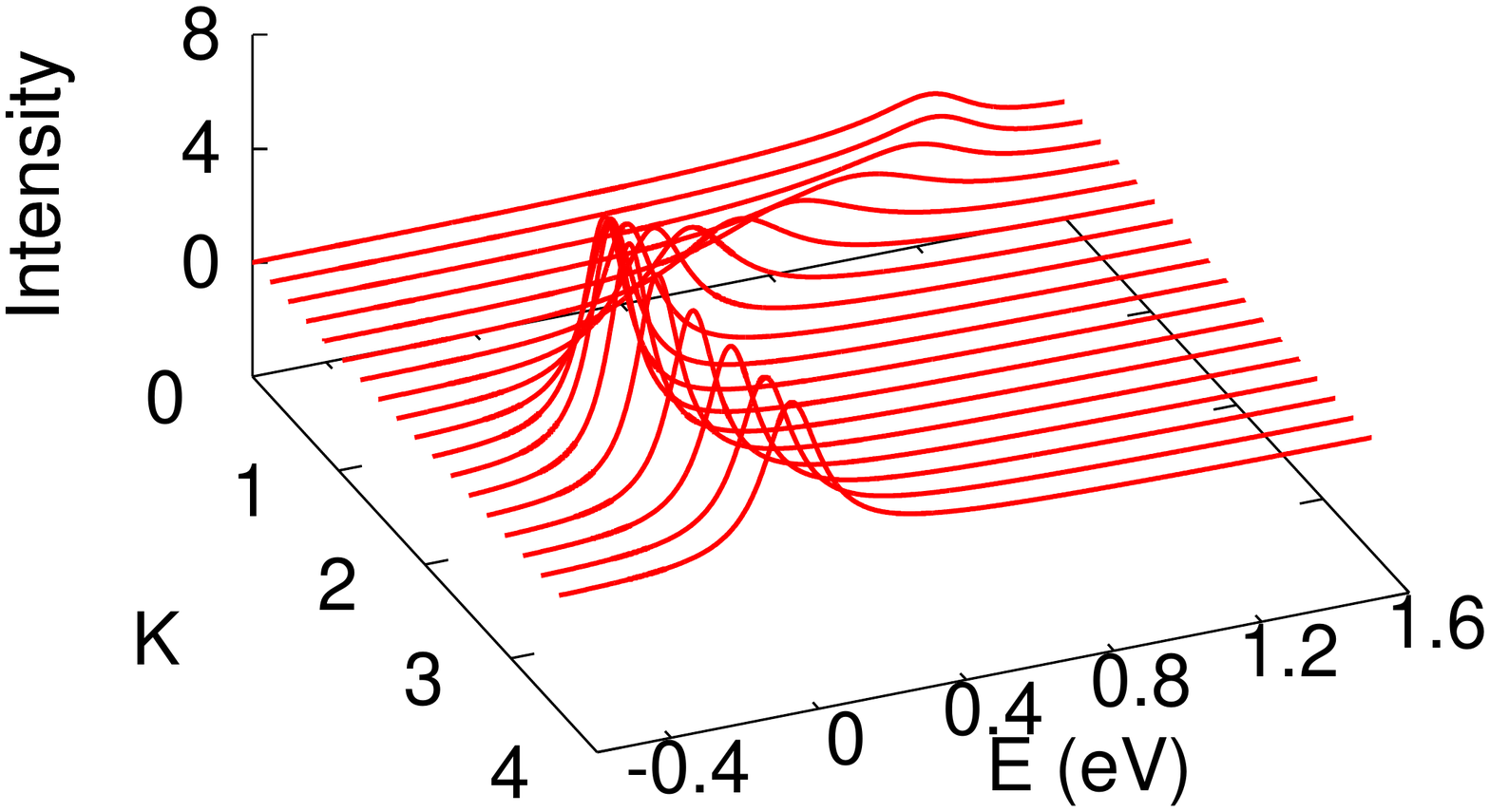}
}
(e)
{\includegraphics[angle=0,width=0.5\columnwidth]{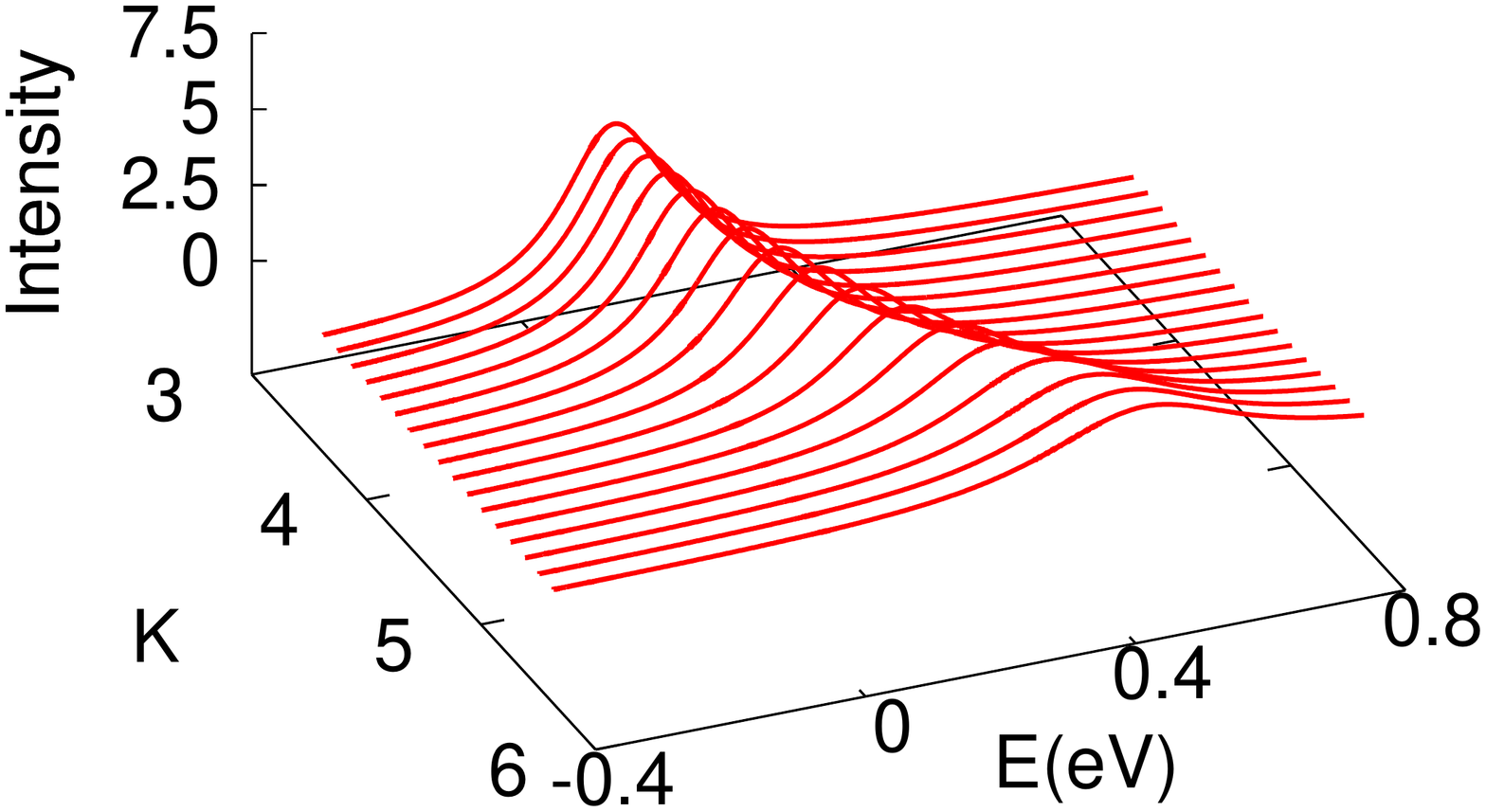}
}
(f)
{\includegraphics[angle=0,width=0.5\columnwidth]{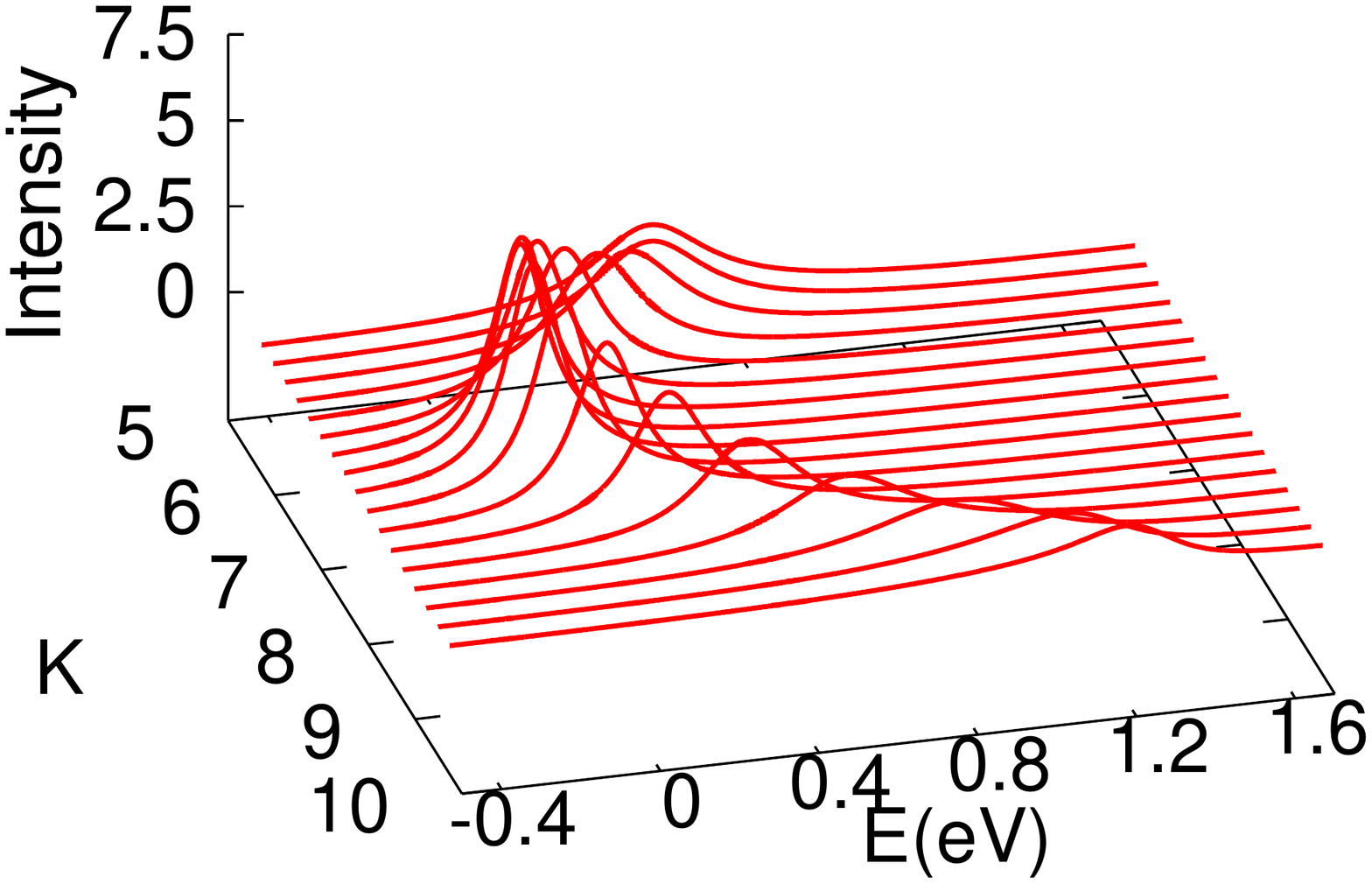}
}
\caption{(Color Online) Calculated ARPES spectra along (i) $\Gamma$-M (ii) M-K 
and (iii) K-$\Gamma$ direction at 20 K (a-c) and 200 K (d-f). $E=0$ marks the Fermi level.}
\label{fig9}
\end{figure*}
\begin{figure}[!ht]
\centering
{\includegraphics[angle=0,width=0.7\columnwidth]{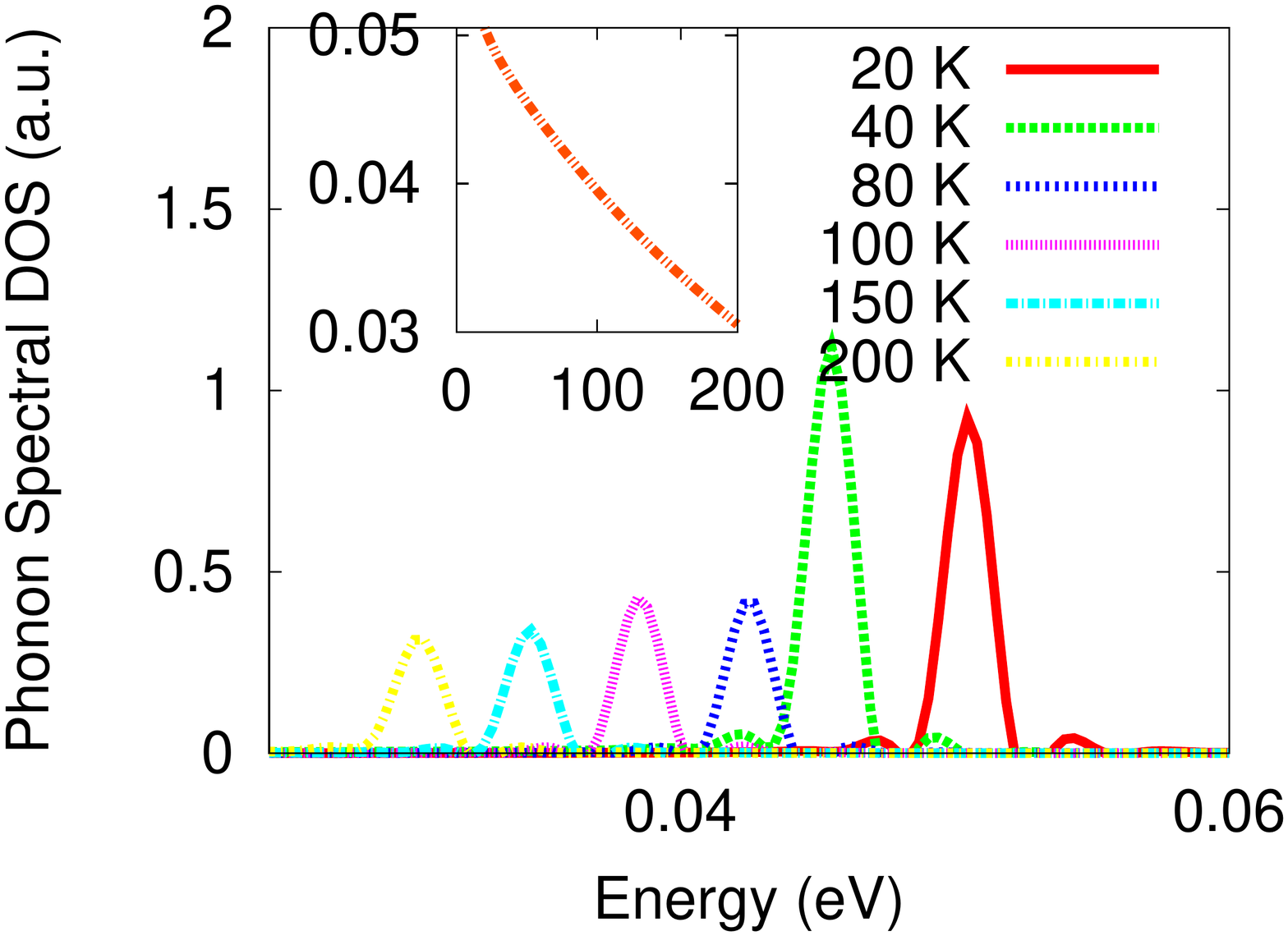}}
\caption{(Color   Online)  Phonon   spectral   density  at   different
  temperatures. The inset shows the peak positions with temperature in
  the normal state.}
\label{fig10}
\end{figure} 
We show how  our theory describes the available  data on 2H-NbSe$_{2}$
without further assumptions. As is well known, DMFT modifies LDA bands
in two  major ways: first,  the static, Hartree terms  renormalize the
relative band  positions (depending on the band  occupations and their
relative   separation),  and   second,   the  dynamical   correlations
generically cause spectral weight transfer (SWT) across a fairly large
energy scale dictated by correlation:  indeed LDA fails in a major way
to incorporate the  physics of this and does not lead  to SWT. We show
the DMFT many-body DOS for Nb-d and Se-p bands in Fig.~\ref{fig2}a and
the  corresponding imaginary and  real parts  of the  self-energies in
Fig.~\ref{fig2}b  and Fig.~\ref{fig2}c  with several  U  and U$_{ab}\,
(t_{ab}=0.1$ eV). The DOS is quite  sensitive to the choice of $U$ and
$U_{ab}$.  Application of small finite U$_{ab}=0.05$ eV for U$=0.7$ eV
causes  instability in  the  Se-$p$ band  at  $\omega=0$, implying  an
orbitally selective insulator with  very poor conductivity (the d-band
has  a small  but finite  DOS  at the  FL). However,  the real  system
remains metallic  throughout, albeit a  poor one, and therefore  U and
U$_{ab}$ are  chosen accordingly,  U$=0.5$ eV,\, U$_{ab}=0.05$  eV, so
that the system remains a poor metal.

However, the real  system is known to show  reasonable phonon coupling
in  its physical properties.  The phonons  do not  appear to  play any
central role in  the unusual normal state properties  or the formation
of CDW~\cite{dord},  they couple with  the excitons and  therefore the
changes in phonon  intensities and lineshape is a  useful tool to shed
light on  the underlying  physics in NbSe$2$.   Our DMFT  results also
indicate that  the phonons  are not responsible  for the  normal state
properties (and  even the  CDW formation, see  later), we  include the
phonons  in  our  calculations   in  the  following  and  study  their
evolution.  In Fig.~\ref{fig3}a we show the resulting many body DOS at
high ($T>T_{cdw}$) and  low T($<T_{cdw}$) for the Nb-d  band for U=0.5
eV  and U$_{ab}=0.05$  eV along  with the  self-energies for  both the
bands. Neglect  of the  mixing term t$_{ab}$  in the  Hamiltonian Eq.1
leads to  results that strongly disagree with  the extant experimental
observations.  Besides, the  excitonic  amplitude will  vanish in  the
ground  state  for t$_{ab}=0$.  Inclusion  of  mixing  produces a  bad
metallic state  with a DOS having  a peak-like structure  at FL, which
goes     down     as    temperature     rises.     We    also     find
Im$\Sigma_{a,b}(\omega=0)>0$  at  $E_F$,  underlining  the  incoherent
normal properties  observed. The  $p$ and $d$  bands continue  to show
renormalized bad  metallic behavior throughout  the temperature range.
Clearly,  this  is caused  by  the  incoherent excitonic  fluctuations
present in the system  already at high temperatures. Quite remarkably,
the self-energy also indicates a  large scattering rate at high $T$ in
the normal  state, reducing as $T$  is lowered - an  indication of the
build up of excitonic coherence  at lower $T$. The peak in self-energy
varies in  an energy range  where phonons are  known to be  active (as
seen by Valla et al.~\cite{valla1}).

If this alternative PEL-based theory is to be valid, then it should be
able  to describe more  experimental observations.  ARPES data  show a
gradual build-up of excitonic correlations with decreasing temperature
as  an  increase  in  the  pseudogap~\cite{boris_prl}.  The  pseudogap
deepens and  the low-energy peak in  PES shifts out  to higher energy,
accompanied by  temperature-induced spectral weight  transfer from low
to high  energy. This is  a real test  for our theory.  It  is already
clear that LDA plus static  HF cannot explain the ARPES lineshapes and
the  spectral weight  transfer.   However, our  formulation must  also
describe the unusual transport properties.  In Fig.~\ref{fig4} we show
our   DMFT  results  for   the  $T$-dependent   optical  conductivity,
$\sigma(\omega,T)$,  calculated  from  the  many body  (DMFT)  Green's
function. Both  the $\omega$ and $T$  dependence of $\sigma(\omega,T)$
are in  agreement with experiment. In  $\sigma(\omega,T)$ a correlated
FL behavior  is never found and  lowering of $T$ shows  a reduction in
the incoherent  scattering. At  low $T$ a  a shoulder-like  feature is
seen around $0.25$~eV. This is  exactly the scale at which a pseudogap
feature  appears  in  the  $T$-dependent  DOS at  low  $T$  (shown  in
Fig.~\ref{fig3}a), strongly  suggesting the tendency  of the preformed
excitons towards  coherence at lower $T$.   Clearly this shoulder-like
feature  in  optical  conductivity  disappears  at  high  $T$  due  to
incoherent  excitonic  effects. DMFT  result  for  $T$ dependent  $dc$
resistivity   is   shown  in   Fig.~\ref{fig5},   which  also   agrees
qualitatively with experimental  data.  The $dc$ resistivity $\rho(T)$
is strongly  dominated by the  incoherent scattering leading to  a non
Fermi  liquid behaviour (nFL);  there is  no $T^2$  regime at  low $T$
i.e., it never shows a correlated Fermi liquid behavior (quite similar
to 2H-TaSe$_2$). The high temperature resistivity is linear in $T$ for
a large  range of $T$ as  observed in the  experiments. The scaterring
rates (\ref{fig5}) track the resistivity and are almost linear at high
$T$,  getting  reduced  as   $T$  decreases  as  in  the  experimental
data~\cite{dord}.  $\rho(T)$  shows a change in slope  at around 30~K,
below which  enhanced metalicity is recovered. We  reconcile this bump
in  resistivity as  coming from  the enhanced  coherence  of preformed
excitons   below  a  crossover   temperature  range   causing  reduced
incoherent scattering. The preformed excitons eventually condense to a
CDW at  low temperatute (see below). This  picture therefore naturally
reconciles with  the near-insensitivity of  the transport data  to the
onset of the commensurate CDW order and enhanced metalicity below it.

The detailed FS map  as a function of $T$ in NbSe$_2$  does not show a
strong  signature of  FS nesting.  The  DMFT FS  can be  approximately
constructed from  the DMFT Green's function~\cite{byc_nat}.  FS map at
lower   temperature,  shown   in  Fig.~\ref{fig6}   in   the  original
unreconstructed   Brillouin  zone,   clearly  shows   that   the  {\it
  correlated}  FS  is  not   affected  significantly  across  the  CDW
transition. The Fermi pockets are visible at $T> T_{CDW}$, albeit with
a  typical scatter  coming from  incoherent scattering  and consequent
loss of quasiparticle feature.  This loss of sharpness correlates well
with the broadness of  lineshape (Fig.~\ref{fig3}) and spectral weight
transfer from  incoherent scattering.  Across $T=T_{CDW}$  there is no
dramatic change: since excitons  are formed at higher temperatures and
parts of the  FS have already been reconstructed  at high $T$, changes
in FS  at the  CDW transition  is therefore small.  This is  of course
corroborated by the  transport data as well. The  normal state is thus
dominated  by  incoherent  scattering  and  the  onset  of  CDW  order
eventually   involves  condensation   of  these   preformed  excitons.
Finally,  given   the  in-plane   normal  state  is   incoherent,  the
out-of-plane  responses  will show  even  more  drastic signatures  of
incoherence~\cite{and-book}.   This  is   borne  out  in  the  optical
conductivity  measurements~\cite{dord}  for  2H-NbSe$_{2}$. Thus,  the
preformed-excitonic  liquid  scenario  provides  a  natural  basis  to
understand  a  wide  variety  of  responses in  the  normal  state  of
2H-NbSe$_{2}$.

Since there  is no prominent or discernible  gap in the FS  at the CDW
transition, Borisenko et al~\cite{boris_prl}. have defined a ``gap" as
the  difference  between  the  binding energies  of  the  leading-edge
midpoints. This  gap is observed in  the normal state also  and is, in
fact, much  larger than  the gap at  the CDW  transition.  Theoretical
ARPES  lineshapes along  K-M-K  direction  and the  gap  are shown  in
Fig.~\ref{fig7}   for  comparison.   Evidently  the   theoretical  gap
structure  tracks  the  experimental  one  fairly  well  as  shown  in
Fig.~\ref{fig7}b. The  gap for temperatures below  T$_{CDW}$ is higher
than the  gap near T$_{CDW}$ Fig.~\ref{fig7}b, suggesting,  in fact, a
minimum  gap  at  the   transition  temperature.  The  CDW  transition
temperature  can thus  be defined  as  the temperature  where the  gap
between  the  leading edge  midpoints  is  a  minimum. The  CDW  order
parameter obtained from the  effective Hamiltonian, is also plotted in
the inset  to Fig.~\ref{fig7}b and it has  the characteristic decrease
with  temperature, although not  following the  static single-particle
mean-field square-root variation.

The evolution  of the exciton-phonon correlation  with temperature can
be     seen     through     the    excitonic     spectral     function
$\rho_{ab}(\omega)=(-1/\pi) $Im$G_{ab}(\omega)$  as well as  the local
excitonic                       average,                      $\langle
(c_{ia\sigma}^{\dag}c_{ib\sigma}+h.c)\rangle=(-1/\pi)
\int_{-\infty}^{\infty}d\omega       Im       G_{ab}(\omega)$       in
Fig.~\ref{fig8}.     The      strong     temperature-dependence     of
$\rho_{ab}(\omega)$  is obvious:  at high  $T$, the  broad, asymmetric
shape is a manifestation of  the incoherent excitonic fluid, while the
low-energy pseudogap  along with  large spectral weight  transfer with
decreasing $T$  signals a progressive build-up  of incipient excitonic
coherence. This is seen more clearly in the $T$-dependence of $\langle
(c_{ia\sigma}^{\dag}c_{ib\sigma}+h.c)\rangle$,   where   a  relatively
steep  increase below  30~K  is visible  (Fig.~\ref{fig8}), while  the
fluctuations  are present till  fairly high  temperatures. Remarkably,
this  exactly   correlates  with  the  broad  bump   in  $\rho(T)$  in
Fig.~\ref{fig5} and is  a clear evidence for the  relevance of strong,
dynamic excitonic  correlations in 2H-NbSe$_{2}$.  A  regime of strong
pair-fluctuation above T$_{cdw}$ and  the existence of a pseudogap far
in to  the normal state where there  is no long range  order, has been
seen  in the  X-ray,  ARPES and  tunnelling  spectroscopy in  NbSe$_2$
recently~\cite{utpal}.  Our results are  indeed quite  consistent with
such observations.   Theoretical ARPES lineshapes  along high symmetry
directions in  Fig.~\ref{fig9} mark exciton-phonon-induced  changes in
the $k$-resolved  spectrum at normal state  and in the  CDW phase. The
ARPES  dispersions match  quite  well with  the  observed spectra  and
present how the incoherent high-temperature normal state is going over
to   a  progressively   coherent  ordered   state,   establishing  the
exciton-phonon  scenario on  a stronger  footing.  Given  the  form of
exciton-phonon  coupling,  by  symmetry,  the  resulting  DMFT  phonon
spectrum (Fig.~\ref{fig10}) should  show particular changes going hand
in hand  with exciton dynamics.  The DMFT phonon spectrum  of NbSe$_2$
shows  maximum intensity  with  reduced linewidth  in  the CDW  phase.
Broadening of  phonon spectrum  above T$_{CDW}$ confirm  coupling with
incoherent excitons at high $T$. As $T$ is lowered below T$_{CDW}$ the
onset  of   excitonic  coherence  reduces   normal  state  scattering,
resulting in  sharpening of the phonon  spectrum. In the  light of our
results, it  is quite  likely that the  instability of  the incoherent
metal to  an UCDW  state at  low $T$ in  2H-NbSe$_{2}$ results  from a
mechanism  where  the one-electron  process  is  incoherent while  the
electronic  correlations  as  well  as the  interactions  mediated  by
exciton-lattice coupling are  relevant. Superconductivity in other TMD
arises  from an  incoherent  metal,  in most  cases  proximate to  CDW
states,  which is  likely to  be the  case here  too. Such  a scenario
appears   to   be  generic   for   all   the   three  TMDs   we   have
studied~\cite{atprl,sktise2} and  may be  applied to other  TMDs also.
Our  approach stresses  on  the central  role  of dynamical  excitonic
correlations  in   small  carrier  density   systems  having  close-by
electron-  and hole-like bands  at $E_F$.  We believe  that approaches
similar to the one presented here and the similarity of the underlying
physics     with     other     TMD     systems     studied     earlier
\cite{atprl,aipp,sktise2}  underline  a common  theme  of  PEL in  the
normal state and its condensation at the CDW tranistion.

With  the  recent  advances  in  the ultrafast  spectroscopy  and  its
application to ordered  electronic states~\cite{hellmann,ruan}, it may
indeed be possible to resolve the issue of a correlation-driven versus
phonon-driven  mechanism for  CDW in  many  systems owing  to the  two
different   time-scales  associated  with   the  two   processes.  The
resolution  of  1T-TiSe$_2$  as  an  electronically  driven  excitonic
mechanism~\cite{hellmann}  lends credence  to  the fact  that the  PEL
mechanism    may     be    operative    in     other    TMD    systems
too~\cite{atprl,sktise2}.   Electronic  Raman  spectroscopy   and  the
determination  of  the nature  and  symmetry  of  the low  temperature
superconducting  states remain  other  experimental observations  that
would shed light on these systems and test the proposed theories.
  
SK acknowledges  CSIR (India) for  financial support through  a senior
research  fellowship.  AT  and  SK acknowledge  useful discussion  and
close  collaboration  on similar  systems  with  Mukul  Laad and  N  S
Vidhyadhiraja. AT acknowledges useful discussions with and inputs from
Chong-Yu Ruan on possible pump probe experiment.


\begin{thebibliography}{59}
\bibitem{aebi} C. Monney, et al., Phys. Rev. B {\bf 81} 155104 (2010).
\bibitem{dord} S. V. Dordevic, et al., Eur. Phys. J. B ~{\bf 33}, 15 (2003).
\bibitem{beck_njp}C. Monney et al., New Journal of Physics~{\bf 12}, 125019 (2010).
\bibitem{peier} R. E. Peierls, {\it Quantum Theory of Solids} (Clarendon, Oxford, 1955).
\bibitem{atprl} A. Taraphder et al., Phys. Rev. Lett {\bf 106}, 236405 (2011).
\bibitem{wilson}J A Wilson, et al., Adv. Phys. {\bf 24}, 117 (1975).
\bibitem{straub} Th. Straub, et al., Phys. Rev. Lett. {\bf 82}, 4504 (1999).
\bibitem{shen} D. W. Shen, et al., Phys. Rev. Lett. {\bf 101}, 226406 (2008).  
\bibitem{johannes08} M. D. Johannes and I.I. Mazin, Phys. Rev. B {\bf 77}, 165135 (2008).
\bibitem{boris_prl} S. V. Borisenko et al., Phys. Rev. Lett {\bf 102}, 166402 (2009).
\bibitem{tonjes} W. C. Tonjes, et al., Phy. Rev. B, {\bf 63}, 235101 (2001). 
\bibitem{moncton} D. E. Moncton et al., Phys. Rev. Lett {\bf 34}, 734 (1975): Phys. Rev. B {\bf 16}, 801 (1977).
\bibitem{murphy} B. M. Murphy et al., Phys. Rev. Lett {\bf 95}, 256104 (2005): J. 
Phys. Condens. Matter {\bf 20}, 224001 (2008).
\bibitem{doran} N. J. Doran et al.,  J. Phys. C: Solid State Phys. {\bf 11}, 685 (1978).
\bibitem{utpal} U. Chatterjee, et al., unpublished and private communication.  
\bibitem{laad} M. Laad, et al., Phys. Rev. Lett.~{\bf 91}, 156402 (2003).
\bibitem{ciuchi} S. Ciuchi et al., Europhys. Lett. {\bf 30}, 151 (1995).
\bibitem{valla1} T. Valla et al., Phys. Rev. Lett {\bf 92}, 086401 (2004). 
\bibitem{and-book} P. W. Anderson, (1997) (Princeton University Press).
\bibitem{byc_nat} K Byczuk et al., Nat. Phys. {\bf 3}, 168 (2007).
\bibitem{aipp} S Koley et al., AIP Conf.Proc. {\bf 1461}, 170 (2012).
\bibitem{sktise2} S Koley et al., arXiv:1212.1026.
\bibitem{note} Iterated Pertubation scheme has been very successfuly used in 
several cases, see for example, V. I. Anisimov, et al. Jour. Phys. Cond. Mat, 
{\bf 9}, 7359 (1997) and M. Laad, et al.~\cite{laad}.
\bibitem{hellmann} S. Hellmann, et al,  Nature Comm.{\bf 3}, 1069 (2012).  
\bibitem{ruan} Tzong-Ru T. Han, et al,  Phys. Rev. B {\bf 86}, 075145 (2012).  
\end{thebibliography}
\end{document}